\documentclass[%
superscriptaddress,
amsmath,
amssymb,
aps,
prb,
floatfix,
]{revtex4-2}

\usepackage{graphicx}
\usepackage{tikz}
\usepackage{multirow}
\usepackage{dcolumn}
\usepackage{bm}
\usepackage{mathtools}
\usepackage{hyperref}
\usepackage{siunitx}
\usepackage[version=4]{mhchem}

\def\njust{Department of Applied Physics, Nanjing University of Science and
Technology, Nanjing 210094, China}
\def\cas{Institute of Physics, Chinese Academy of Sciences, Beijing 100190,
China}
\def\ucas{School of Physical Sciences, University of Chinese Academy of Sciences, Beijing 100049, China}
\def\ustc{University of Science and Technology of China, Hefei 230026, China}
\def\aisi{AI for Science Institute, Beijing 100080, P. R. China}
\def\pku{HEDPS, CAPT, School of Mechanics and Engineering Science 
    and School of Physics, Peking University, Beijing, 100871, P. R. China}

\newcommand{\diag}{\operatorname{diag}}
\newcommand{\sym}{\operatorname{sym}}
\newcommand{\xc}{\mathrm{xc}}

\def\bd{{\mathsf{d}}}
\def\tr{\operatorname{tr}} 
\def\sym{\operatorname{sym}} 
\def\grad{\operatorname{grad}} 

\def\bg{{\mathsf{g}}}
\def\bk{{\mathbf{k}}}

\def\br{{\mathbf{r}}}

\def\bv{{\mathsf{v}}}
\def\bx{{\mathsf{x}}}

\def\bG{{\mathbf{G}}}

\def\dd{{\mathop{}\!\textrm{d}}}
\def\dr{{\dd\br\,}}

\def\H{{\textrm{H}}}

\def\x{{\textrm{x}}}
\def\xc{{\textrm{xc}}}

\def\half{\frac{1}{2}}

\begin{document}

\title{Iterative minimization in reduced density matrix functional theory for periodic systems}

\author{Kai Luo}
\email{kluo@njust.edu.cn}
\affiliation{\njust}

\author{Jingang Han}
\affiliation{\cas}
\affiliation{\ucas}


\author{Peize Lin}
\affiliation{\ustc}

\author{Daye Zheng}
\affiliation{\aisi}

\author{Mohan Chen}
\affiliation{\pku}

\author{Xinguo Ren}
\email{renxg@iphy.ac.cn}
\affiliation{\cas}

\date{prepared on Jul 30, 2026}

\begin{abstract}
Reduced density matrix functional theory (RDMFT) offers a route beyond Kohn-Sham
density functional theory
for strongly correlated systems, yet practical calculations for periodic solids are still out of reach.
We formulate RDMFT for extended systems in a basis-independent way and present a planewave implementation 
using iterative minimization for periodic solids, evaluating nonlocal exchange-correlation functionals
through the existing adaptive compressed exchange machinery.
Natural occupations are optimized under N-representability constraints
with a spectral projected gradient (SPG) method or an first-order explicit-by-implicit
(EBI) map, while natural orbitals are updated by Riemannian optimization on the complex Stiefel manifolds.
Benchmarks on typical systems of \ce{H2}, silicon, and sodium with the Hartree-Fock functional show that SPG
reproduces converged hybrid references, whereas EBI can stall when occupations approach $0$ or $1$. With
the power and Müller functionals, SPG yields lower energies and more stable convergence than EBI.
Applications to fractionally charged \ce{LiH}, dissociating \ce{H2} and \ce{N2} molecules, and equation of state of silicon
show that the algorithm presented in this implementation is reliable and robust.  
\end{abstract}

\maketitle

\section{Introduction}
\label{sec:intro}
For decades, the Kohn-Sham density functional theory (DFT) \cite{Hohenberg1964,Kohn1965} has been the most widely used method 
in chemistry and materials science for finite and extended systems. However, it faces 
challenges in handling systems where the delocalization errors and strong correlations play a central role. 
Reduced density matrix functional theory (RDMFT) \cite{Gilbert1975,Muller1984} provides a promising alternative \textit{ab initio} approach 
to overcome the inherent limitations of DFT by going beyond the idempotency constraint on the one-body reduced density matrix (1-RDM). 
Gilbert's extension of the Hohenberg-Kohn theorem to nonlocal external potentials provides
the formal basis for treating the ground-state energy as a functional of the one-body reduced density matrix (1-RDM)
\cite{Gilbert1975}. Because the 1-RDM
retains orbital information and admits non-idempotent spectra, it is a natural
language for bond dissociation, near degeneracy, and correlated insulating
behavior \cite{Pernal2016}.

The practical performance of RDMFT depends on two key factors: the accuracy of the approximation chosen for the
exchange-correlation (XC) functional and the efficiency and robustness of the solver for finding the density matrix solution.
Early and still influential XC functionals include M\"uller's functional \cite{Muller1984} (or the BB functional \cite{BB2002}), 
the Goedecker-Umrigar self-interaction-corrected variant \cite{Goedecker1998}.
For molecules, a family of BB-corrected functionals to fix BB's overcorrelation problem have been developed based on 
notion of strong and weak occupation \cite{Gritsenko2005}.
Interestingly, Csányi and collaborators constructed CHF and CGA functionals from a tensor-product ansatz\cite{Csanyi2000,Csanyi2002}.
Piris and collaborators developed another family of natural-orbital functionals,
known as PNOF$i$ ($i=1$--$7$), by approximating the cumulant matrix subject to known constraints on the two-body reduced density matrix $\Gamma$.
Pernal established a correspondence between the PNOF5 functional and the antisymmetrized product of strongly orthogonal geminals \cite{Pernal2013}.
Motivated by range separation, related hybrids combine a density-functional approximation at short range with the long-range M\"uller functional \cite{Pernal2010,Ai2023}.
For periodic solids, partly due to the lack of packages, there are fewer functionals and their performance is not well studied. So far, the power functional introduced by Sharma \textit{et al.}
extended RDMFT beyond molecules and atoms, showing that fractional powers of the
1-RDM can reproduce band gaps in ordinary semiconductors and open correlated gaps in Mott insulators\cite{Sharma2008}. 
The improved CGA functional was meant to treat solids by matching the uniform electron gas (UEG) correlation, but its performance is still unknown.
Further work aligning with UEG correlation by Baldsiefen \textit{et al.} at zero temperature led to a few functionals,
but no systematic study has been done yet \cite{Baldsiefen2017}.

The other key factor is the numerical bottleneck for solving the density matrix equation. Recently,
Vladaj \textit{et al.} proposed a variational minimization scheme for functionals of the 1-RDM in the ensemble N-representable domain to  solve the density matrix equation 
\cite{Vladaj2024}, which seems quite efficient on the Hubbard model and hydrogen molecule. A more general approach is necessary 
to include versatile functionals that is not restricted to 1-RDM only functionals.
Following the conventional 
spectral decomposition, the density matrix is represented by the natural orbitals and their associated natural occupation numbers.
Natural occupations must satisfy N-representability, orbitals must remain orthonormal in either Euclidean
or overlap-weighted metrics, and the total energy is generally nonconvex in the
combined variables.
These issues become more pronounced in periodic
calculations, where spin channels, $k$-point weights, and nonlocal exchange-like
operators must be handled together. Efficient implementations therefore require
more than a formal XC functional: they need stable constraint handling,
geometry-aware orbital optimization, and high performance exact-exchange (EXX) evaluation in a parallel layout that can serve both
molecules and solids. 
Most implementations for RDMFT calculations are based on existing 
quantum chemistry packages, which are mostly oriented for molecules and clusters. 
It is worth noting that an open-source solution provided  by DoNOF \cite{DoNOF2021,DoNOF2026} sets a good example for accessible algorithms for RDMFT development.
So far, as far as we know, only the LAPW package \textsc{elk} \cite{elk} can perform such calculations 
for solids. There these natural orbitals are expanded in a set of converged KS orbitals
and EXX matrix elements are stored locally on the hard-disk. 
Therefore, the implementation lacks the flexibility and efficiency for general applications.

Here we provide a basis-independent formulation of RDMFT for periodic solids
and implement it as a self-consistent-field module in a massively parallel planewave code, 
Quantum ESPRESSO (QE)\cite{Giannozzi2009}.
 While natural orbitals are updated by Riemannian optimization on complex Stiefel manifolds following our previous work \cite{Luo2025}, 
 natural occupations are optimized under N-representability constraints with the spectral projected gradient (SPG) method. 
 The explicit-by-implicit (EBI) map of first-order is included for comparison.
We evaluate nonlocal XC functionals through the adaptive compressed exchange (ACE) \cite{Lin2016} 
framework that is already used in hybrid calculations.
Representative benchmarks on molecules and bulk (\ce{H2}, silicon, and sodium)
show that this workflow is numerically stable and suitable for general
solid-state applications.

The remainder of the paper is organized as follows.
Section~\ref{sec:theory} develops a basis-independent RDMFT formulation for
periodic systems, including the $k$-resolved spectral
representation of the 1-RDM, XC functionals and their
evaluation, the constrained energy minimization problem, and the 
alternating updates for natural occupations via SPG or EBI  and for natural orbitals
on complex Stiefel manifolds.
Section~\ref{sec:computational_details} describes the planewave
implementation in QE, where XC
functionals are evaluated through ACE and analytical gradients are verified
against finite differences.
Section~\ref{sec:results} presents Hartree-Fock (HF) and power functional
benchmarks of SPG against EBI, tests of initial-occupation robustness, and
applications to fractionally charged \ce{LiH}, molecular dissociation, and the
silicon equation of state.
Section~\ref{sec:summary} concludes with a summary and outlook.

\section{Theory} 
\label{sec:theory}

\subsection{Reduced density matrix functional theory}
\label{sub:rdmft}
We consider the systems of interacting electrons governed by the Hamiltonian
\begin{equation}
    \hat{H} = \hat{T} + \hat{V} + \hat{W} ,
\end{equation}
with the kinetic energy $\hat{T}$, 
the external one-body potential $\hat{V}$, and the two-body electron-electron interaction $\hat{W}$, respectively.
The ground-state energy is achieved by variationally minimizing the Hamiltonian with respect to the 
wave function $\Psi$.
In RDMFT, the ground-state energy can be obtained by minimizing the 
energy $E[\gamma]$ as a functional of 1-RDM, $\gamma(x,x')$,
 for fixed number of electrons $N$. Here $x$ is a composite of the spatial and spin coordinates, $x = (\br, \sigma)$.
 Following the conventional decomposition of the energy, we may write 
 \begin{equation}
    E[\gamma] = E_k[\gamma] +E_{\H}[\gamma]+E_{\xc}[\gamma]  + V[\gamma].
 \end{equation}
 These functionals of kinetic energy, Hartree energy, and external potential energy are
 \begin{eqnarray}
 E_k[\gamma] &=& \int d x' \lim _{\br \rightarrow \br'}\left(-\frac{\nabla^2}{2}\right) \gamma\left(x, x'\right), \\
 E_{\H}[\gamma] &=& \frac{1}{2} \int d x d x' w\left(x, x'\right) \gamma(x, x) \gamma\left(x', x'\right), \\
 V[\gamma] &=& \int d x \int d x' v\left(x, x'\right) \gamma\left(x', x\right),
\end{eqnarray}
where $w(\br,\br')$ is the electron-electron interaction potential (e.g. $1/|\br-\br'|$ for Coulomb interaction) 
and the external potential $v(x,x')$ is kept nonlocal in general. 
At finite temperature $T$, the free energy $F[\gamma]$  is given by
adding the entropic energy, $F[\gamma] =  E[\gamma] - T S[\gamma]$. The entropy can be computed from the 
natural occupation numbers $n_i$,
\begin{equation}
    S[\gamma] = - k_B \sum_i\left[n_i \ln n_i+\left(1-n_i\right) \ln \left(1-n_i\right)\right],
\end{equation}
where $k_B$ is the Boltzmann constant. This formula also defines the XC energy, 
which should include the difference of the entropy term between the interacting and 
non-interacting systems. The exact form of XC energy $E_\xc[\gamma]$ is unknown and 
has to be approximated; Sec.~\ref{sub:xc} summarizes the XC
functionals used in this work.

Diagonalization of the Hermitian $\gamma(x, x')$, yields the natural occupation numbers $n_i$ and natural orbitals $\psi_i(x)$,
which gives the spectral representation of the 1-RDM \cite{Lowdin1955},
\begin{equation}
    \gamma(x, x') = \sum_i  n_i \psi_i(x) \psi_i^*(x').
    \label{eq:spectral_representation}
\end{equation}
The N-representability condition for 1-RDM was solved by Coleman \cite{Coleman1961,Coleman1963}.
The theorem states that the 1-RDM is N-representable if and only if the natural occupation
numbers are in the interval [0,1] and the sum of the natural occupation numbers is equal to the number of electrons.
Mathematically, it can be summarized as the following constraints:
\begin{equation}
    \sum_i^{N_\mathrm{b}}  n_i = N, \quad 0 \le n_i \le 1,
    \label{eq:occ_constraint}
\end{equation}
where $N_\mathrm{b}$ is the number of occupied natural orbitals and $N$ is the number of electrons. For 
the cuspy Coulomb interaction at zero temperature, an infinite number of natural orbitals in principle is required 
to fully represent the non-idempotent 1-RDM, namely $N_\mathrm{b}\to \infty$. In practice, 
$N_\mathrm{b}$ is typically limited by the number of basis functions.
Often, $N_\mathrm{b}$ could be further limited to a minimum number, especially for 
these idempotent density matrices in HF theory or DFT.

Under the periodic boundary conditions, $\gamma$ can be written as a sum of $\bk$-resolved 1-RDMs,
\begin{equation}
\gamma(\x,\x') = \sum_{\bk} w_{\bk} \gamma_{\bk}(\x,\x') \,.
\label{eq:kresolved_rdm}
\end{equation}
In the collinear spin-polarized case, the $\bk$-resolved 1-RDM $\gamma_{\bk}(\x,\x')$ is diagonal in spin,
and it can be written in the spectral form as
\begin{equation}
\gamma_{\bk}(\x,\x') =  \delta_{\sigma\sigma'} \sum_{i=1}^{N_\mathrm{b}}  n_{i\bk \sigma} \, \psi_{i\bk}(\x) \,
\psi^{*}_{i\bk}(\x').
\label{eq:gamma_spectral}
\end{equation}
 These natural orbitals $\psi_{i\bk}$ follow the Bloch theorem by the translational symmetry of the 
 $\bk$-resolved 1-RDM.
For periodic systems, $w_{\bk}$ are Brillouin-zone weights normalized so that $\sum_{\bk} w_{\bk} = 1$ 
in a spin-polarized calculation 
(or $2$ in the usual spin-restricted closed-shell
convention where the trace over spin is absorbed into weights).
For collinear spin-polarized calculations, the spin densities $\rho_{\sigma}(\br)=\gamma(\br\sigma,\br\sigma)$
($\sigma=\uparrow,\downarrow$) sum to
$\rho(\br)=\rho_{\uparrow}(\br)+\rho_{\downarrow}(\br)$, and collinear magnetization profiles
use $m_z(\br)=\rho_{\uparrow}(\br)-\rho_{\downarrow}(\br)$.
In addition to the electron number constraint $\sum_{i \bk\sigma} w_{\bk} n_{i\bk\sigma}=N_e$, 
a fixed magnetization $M_z$ can be enforced with an additional Lagrange multiplier on 
$\sum_{i \bk} w_{\bk} (n_{i\bk\uparrow}-n_{i\bk\downarrow}) - M_z$.
Isolated systems such as atoms and molecules correspond to the $\Gamma$-only limit, and $w_{\Gamma}=1$, so
Eq.~\eqref{eq:kresolved_rdm} covers both finite and extended systems without
changing notation.

These Bloch natural orbitals $\psi_{i\bk}$ are mutually orthogonal within each
$\bk$ block [Eq.~\eqref{eq:orb_constraint}] and are normally expanded in a Bloch basis $\{ \chi_{\mu\bk} \}$,
\begin{equation}
\psi_{i\bk}(\br) = \sum_{\mu=1}^{N_{\mathrm{basis}}} C_{\mu i \bk} \, \chi_{\mu\bk}(\br)\,.
\label{eq:lcao_expansion}
\end{equation}
The dimension of the basis set is $N_{\mathrm{basis}}$, with $N_{\mathrm{basis}} \ge N_\mathrm{b}$.
The expansion coefficients $C_{\mu i \bk}$ can be regarded as a matrix whose $i$th column
contains the expansion coefficients of the $i$th natural orbital indexed by $\bk$.
Its dimension is $N_{\mathrm{basis}} \times N_\mathrm{b}$ and it is shortly denoted as $\mathsf{C}_{\bk}$. 

In general, the overlap matrix for orthogonal basis (e.g. planewave basis) is
simply the identity matrix $\mathsf{I}_{N_{\mathrm{basis}}}$.
Orthonormality of natural orbitals implies
\begin{equation}
(\mathsf{C}_{\bk})^\dagger \mathsf{S}_{\bk} \mathsf{C}_{\bk} = \mathsf{I}_{N_\mathrm{b}},
\label{eq:orb_constraint}
\end{equation}
with overlap matrix $(\mathsf{S}_{\bk})_{\mu\nu} \equiv \langle \chi_{\mu\bk} | \chi_{\nu\bk} \rangle$.
Collecting the occupation numbers $n_{i\bk}$ into a vector $\mathsf{n}_{\bk}$,
the 1-RDM in the basis set $\{ \chi_{\mu\bk} \}$ can be compactly written in the matrix form as
\begin{equation}
    \gamma_{\bk} = \mathsf{C}_{\bk} \, \diag(\mathsf{n}_{\bk}) \, \mathsf{C}_{\bk}^\dagger,
\label{eq:1rdm}
\end{equation}
when $\diag(\bv)$ is a diagonal matrix with elements $v_i \delta_{ij}$.
The electron density is given by the diagonal of the 1-RDM,
\begin{equation}
\rho(x) = \gamma(x,x')\big|_{x'=x} =\sum_{\bk} w_{\bk}
\sum_i n_{i\bk\sigma} \rho_{i\bk}(x) \,,
\end{equation}
where each orbital density $\rho_{i\bk}(x)$ can be obtained from these coefficient matrices $\mathsf{C}_{\bk}$ via
\begin{equation}
    \rho_{i\bk}(x) =  |\psi_{i\bk}(x)|^2 = \sum_{\mu, \nu}^{N_{\mathrm{basis}}}  C_{\mu i \bk} C_{ \nu i \bk}^*
    \chi_{\mu\bk}(x) \chi_{\nu\bk}^*(x).
\label{eq:orbital_density}
\end{equation}

\subsection{Exchange-correlation functionals}
\label{sub:xc}
The XC energy $E_{\xc}[\gamma]$ is a functional of the full
1-RDM rather than of the density alone.
In the natural-orbital representation,
\begin{equation}
E_{\xc}
= - \frac{1}{2} \sum_{\bk,\bk'} w_{\bk}\, w_{\bk'} \sum_{ij}
f(n_{i\bk},n_{j\bk'})
\iint \frac{\psi_{i\bk}^*(\br)\, \psi_{j\bk'}^*(\br')\,
\psi_{i\bk}(\br')\, \psi_{j\bk'}(\br)}{|\br-\br'|}\,
\mathrm{d}\br\, \mathrm{d}\br',
\label{eq:e_xc_general}
\end{equation}
where $f(n_i,n_j)$ is a pair kernel acting on natural-occupation pairs. 
Table~\ref{tab:functionals} lists commonly used XC
functionals and their pair kernels $f(n_i,n_j)$.
\begin{table}[t]
\caption{XC couplings used in the present formulation.
For separable forms, $f(n_i,n_j)=g(n_i)\,g(n_j)$.
For the power functional, $\alpha\in(\tfrac{1}{2},1)$; values
$\alpha\approx 0.656$ (periodic semiconductors) and
$\alpha\approx 0.525$ (molecular dissociation) appear in the literature
\cite{Sharma2008}.
Mixed functionals whose terms are each of the form $g(n_i)\,g(n_j)$
are separable and evaluated as sums of ACE channels.}
\label{tab:functionals}
\begin{ruledtabular}
\begin{tabular}{lll}
Functional & $f(n_i,n_j)$ & Separability / channels \\
\hline
HF & $n_i n_j$ & Separable; $g(n)=n$ \\
M\"uller \cite{Muller1984} & $\sqrt{n_i n_j}$ & Separable; $g(n)=n^{1/2}$ \\
Power \cite{Sharma2008} & $n_i^{\alpha} n_j^{\alpha}$ & Separable; $g(n)=n^{\alpha}$ \\
Goedecker-Umrigar \cite{Goedecker1998} & $\sqrt{n_i n_j}$ for $i\neq j$,
$n_i^2$ for $i=j$ & Separable \\
CHF \cite{Csanyi2000} & $\tfrac{1}{2} n_i n_j +
\tfrac{1}{2}\sqrt{n_i(1-n_i)}\sqrt{n_j(1-n_j)}$ & Separable; 2 channels \\
CGA \cite{Csanyi2002} & $\tfrac{1}{4} n_i n_j +
\tfrac{1}{4}\sqrt{n_i(2-n_i)}\sqrt{n_j(2-n_j)}$ & Separable; 2 channels \\
BOW \cite{Baldsiefen2017} & $(n_i n_j)^{\alpha}-\alpha n_i n_j +
\alpha -\alpha(1-n_i n_j)^{1/\alpha}$, $\alpha=0.61$ & Non-separable \\
\end{tabular}
\end{ruledtabular}
\end{table}
When the pair kernel can be written as a sum of terms, each of which factorises as
\begin{equation}
f(n_{i\bk},n_{j\bk'}) = g(n_{i\bk})\, g(n_{j\bk'}),
\label{eq:xc_separable}
\end{equation}
it is said to be \emph{separable}, with single-occupation coupling function
$g(n)$.
In this work we restrict to separable functionals; non-separable kernels such
as the BOW functional \cite{Baldsiefen2017} require the full pair and are
substantially more costly to evaluate.

In a planewave basis, Eq.~\eqref{eq:e_xc_general} with the separable form of
Eq.~\eqref{eq:xc_separable} becomes
\begin{equation}
E_{\xc}
= - \frac{1}{2} \sum_{\bk,\bk'} w_{\bk}\, w_{\bk'} \sum_{ij}
g(n_{i\bk})\, g(n_{j\bk'}) \sum_{\bG}
\frac{4\pi}{|\bG-\bk+\bk'|^{2}}\,
\mathcal{Y}_{i\bk,j\bk'}(\bG)\,
\mathcal{Y}_{j\bk',i\bk}(-\bG),
\label{eq:exx_gspace}
\end{equation}
where $\mathcal{Y}_{i\bk,j\bk'}(\bG)$ are the Fourier transforms of the
codensities $\psi_{i\bk}^*(\br)\psi_{j\bk'}(\br)$,
\begin{equation}
\mathcal{Y}_{i\bk,j\bk'}(\bG)
= \int_{\Omega} e^{-i(\bG-\bk+\bk')\cdot\br}\,
\psi_{i\bk}^*(\br)\, \psi_{j\bk'}(\br)\,\mathrm{d}\br,
\label{eq:codensity_ft}
\end{equation}
and the $\bG=\bk-\bk'$ Coulomb singularity is regularised by the
Gygi-Baldereschi prescription \cite{Gygi1986}.
As noted above, $E_{\xc}$ is evaluated through ACE.

\subsection{Iterative minimization}
\label{sub:minimization}
For a fixed $N$ electrons, the energy solution can be found by minimizing the objective 
function $E$ 
\begin{equation}
   E_{\mathrm{gs}}=  \displaystyle\min_{\{\mathsf{n}_{\bk}\},\, \{\mathsf{C}_{\bk}\} } E [\{\mathsf{n}_{\bk}\}, \{\mathsf{C}_{\bk}\}] \\
\end{equation}
subject  to the N-representability constraints Eq.~\eqref{eq:occ_constraint}
 and the orthogonality constraint Eq.~\eqref{eq:orb_constraint}, namely,
\begin{equation}
\left\{
\begin{aligned}
    \sum_{i \bk} w_{\bk} n_{i\bk} - N_e &= 0, \\
    0 \le n_{i\bk}\le 1  &, \\
    (\mathsf{C}_{\bk})^\dagger \mathsf{S}_{\bk} \mathsf{C}_{\bk} - \mathsf{I}_{N_\mathrm{b}} &= 0 .
\end{aligned}
\right.
\end{equation}
for all $\bk$.
The objective function is  
\begin{eqnarray}
    E[\gamma]  &=&   \sum_{i \bk} w_{\bk} n_{i\bk}
    \int \dr \psi_{i\bk}^* (\br) h_{\bk} \psi_{i\bk} (\br) 
    + \frac{1}{2} \iint d\br \, d\br'  \frac{\rho(\br)\rho(\br')}{|\br-\br'|} + E_\xc[\gamma],
    \label{eq:energy_functional}
\end{eqnarray}
where one-body Hamiltonian $h_{\bk} = T_{\bk} + V^{\mathrm{loc}}_{\bk} +
V^{\mathrm{nl}}_{\bk}$ collects kinetic, local pseudopotential, and
nonlocal pseudopotential contributions in the usual Kohn-Sham sense, and
$E_{\xc}[\gamma]$ is given by Eq.~\eqref{eq:e_xc_general} with separable
kernels as in Eq.~\eqref{eq:xc_separable}.

The solution of the minimization problem gives the 
 ground-state 1-RDM $\gamma_\mathrm{gs}$ in the spectral representation.
To find the solution, we adopt the iterative minimization approach by updating the occupation numbers and the orbital coefficients in an alternating fashion in each iteration.
This approach requires the gradient of the objective function with respect to the occupation numbers and the orbital coefficients,
\begin{equation}
    \nabla_{\mathsf{n}_{\bk}} E(\{\mathsf{n}_{\bk}\}, \{\mathsf{C}_{\bk}\}), \quad \text{and} \quad
    \nabla_{\mathsf{C}_{\bk}} E(\{\mathsf{n}_{\bk}\}, \{\mathsf{C}_{\bk}\}).
\end{equation}

\subsubsection{Occupation number optimization}
\label{subsub:occ_opt}
The occupation numbers obey two constraints in Eq.~\eqref{eq:occ_constraint}. 
One is the electron-number constraint, and the other is box constraint. 
Previously for the box constraint,  the constrained optimization problem is converted into an unconstrained problem,
by introducing a  parameterization function. Two commonly used functions are
the cosine square function $\cos^2 \theta$ and the sigmoid function $\frac{1}{1+e^{-x}}$.

The electron-number constraint can then be
handled by using a  Lagrangian multiplier. An improved treatment is to add a quadratic penalty term to the objective function,
which is known as the augmented Lagrangian method.
With  $c(\mathsf{n}) = \sum_{i \bk} w_{\bk} n_{i\bk} -N_e$, the ALM objective function $E_{\textrm{ALM}}$ is
\begin{equation}
E_{\textrm{ALM}}(\mathsf{n},\lambda,\mu)
= E(\mathsf{n}) + \lambda c(\mathsf{n}) + \frac{\mu}{2} c(\mathsf{n})^2,
\label{eq:augmented_lagrangian}
\end{equation}
where parameters $\lambda$ and $\mu$ are parameters to be determined. 
The Lagrange multiplier are updated, $\lambda \leftarrow \lambda + \mu\,
c(\mathsf{n})$, based on the Karush-Kuhn-Tucker (KKT) optimality conditions \cite{Kuhn2014}, while the penalty parameter $\mu$ is adjusted heuristically to improve convergence, e.g. $\mu$ may be increased by a factor about 2-5 if $|c|$ remains large.
This ALM method typically requires large number of iterations 
to converge and is less optimal when evaluation of 
the objective function is expensive. The electron-number constraint is also only approximately achieved when convergence is reached.

The earliest implementation for solid is probably in the LAPW code \textsc{ELK} by Sharma et al.~\cite{Sharma2008}.
There, the occupation numbers are optimised directly in $\mathsf{n}$-space on a straight,
box-respecting segment that preserves the electron count exactly. Recently, the usage of EBI map for the molecular systems 
has been proposed by Yao \textit{et al.}~\cite{Yao2021,Yao2022}. EBI replaces the box-constrained vector $\mathsf{n}$ by
 unconstrained parameters $\bx$ and a scalar parameter $\mu$ that enforces
 the electron count implicitly. Instead of using the Newton's method, we choose to use the gradient-based method,
 which only needs the first-order gradient. Per band $i$ and  point $\bk$,
 \begin{equation}
 n_{i\bk} = \tfrac{\mathrm{erf}(x_{i\bk} + \mu) + 1}{2},
 \label{eq:ebi_map}
 \end{equation}
 with $\mu$ found by bisection on
 $\zeta(\mu)=\sum_{i\bk} w_{\bk} n_{i\bk}-N_e=0$.
Writing $u_{i\bk}\equiv x_{i\bk}+\mu$, the chain rule gives
\begin{equation}
\frac{\partial n_{i\bk}}{\partial u_{i\bk}}
= \tfrac{1}{2}\,\mathrm{erf}'(u_{i\bk})
= \frac{1}{\sqrt{\pi}}\, e^{-u_{i\bk}^2},
\label{eq:ebi_dn_du}
\end{equation}
using $\mathrm{erf}'(x) = \frac{2}{\sqrt{\pi}} e^{-x^2}$.
Following the EBI supplementary derivation~\cite{Yao2021}, $\mu$ is an
implicit function of $\bx$ fixed by the electron-count constraint
$\zeta(\bx,\mu)=\sum_{i\bk} w_{\bk} n_{i\bk}-N_e=0$.
For any fixed $\bx$, $\zeta$ is strictly monotone in $\mu$ because
$\partial n_{i\bk}/\partial u_{i\bk}>0$ and $w_{\bk}>0$, so a unique root
$\mu(\bx)$ exists.
Differentiating $\zeta(\bx,\mu(\bx))=0$ with respect to $x_{i\bk}$ then gives
\begin{equation}
\sum_{j\bk'} w_{\bk'}\,
\frac{\partial n_{j\bk'}}{\partial x_{i\bk}} = 0.
\label{eq:ebi_dN}
\end{equation}
With $u_{j\bk'}=x_{j\bk'}+\mu$ and the chain rule,
\begin{equation}
\frac{\partial n_{j\bk'}}{\partial x_{i\bk}}
=
\frac{\partial n_{j\bk'}}{\partial u_{j\bk'}}
\left(
\delta_{ij}\delta_{\bk\bk'}
+ \frac{\partial \mu}{\partial x_{i\bk}}
\right),
\label{eq:ebi_dn_dx}
\end{equation}
Eq.~\eqref{eq:ebi_dN} becomes
\begin{equation}
w_{\bk}\,\frac{\partial n_{i\bk}}{\partial u_{i\bk}}
+ \frac{\partial \mu}{\partial x_{i\bk}} \sum_{j\bk'} w_{\bk'}\,
\frac{\partial n_{j\bk'}}{\partial u_{j\bk'}} = 0,
\qquad\text{hence}\qquad
\frac{\partial \mu}{\partial x_{i\bk}}
= -\frac{w_{\bk}\, \frac{\partial n_{i\bk}}{\partial u_{i\bk}}}
       {\sum_{j\bk'} w_{\bk'}\, \frac{\partial n_{j\bk'}}{\partial u_{j\bk'}}},
\label{eq:ebi_dmu}
\end{equation}
where the denominator is $\partial\zeta/\partial\mu\neq 0$.
(In the molecular EBI formulae the Brillouin-zone weights are unity and the
sums run over orbitals only~\cite{Yao2021,Yao2022}.)
With the occupation gradient $g_{i\bk}= \partial E/\partial n_{i\bk}$,
the gradient with respect to $x_{i\bk}$ follows from the chain rule
$\partial E/\partial x_{i\bk}
= \sum_{j\bk'} g_{j\bk'}\, \partial n_{j\bk'}/\partial x_{i\bk}$,
which leads to
\begin{equation}
 \frac{\partial E}{\partial x_{i\bk}}
 = \frac{\partial n_{i\bk}}{\partial u_{i\bk}}
 \left(g_{i\bk} - w_{\bk}\, R\right),
 \qquad
 R \equiv \frac{\sum_{j\bk'} w_{\bk'}\, g_{j\bk'}\,
               \frac{\partial n_{j\bk'}}{\partial u_{j\bk'}}}
 {\sum_{j\bk'} w_{\bk'}\, \frac{\partial n_{j\bk'}}{\partial u_{j\bk'}}}.
 \label{eq:ebi_grad}
 \end{equation}
The variable $x_{i\bk}$ is updated with $x_{i\bk} \leftarrow x_{i\bk} - \tau   \frac{\partial E}{\partial x_{i\bk}} $
where step size of $\tau$ is found by a simple backtracking Armijo line search.

In this work, we minimize $E(\mathsf{n})$ at fixed natural orbitals using the
spectral projected gradient method (SPG) of Birgin \textit{et
al.}~\cite{Birgin2000}.
For a periodic solid, every trial occupation vector must satisfy the box
bounds $0\le n_{i\bk}\le 1$ and the BZ-weighted electron-count
constraint $\sum_{i\bk} w_{\bk} n_{i\bk}=N_e$.
We collect all such vectors in the set
\begin{equation}
\Omega =\Bigl\{\mathsf{n}:\ 0\le n_{i\bk}\le 1,\ \sum_{i \bk}
w_{\bk} n_{i\bk}=N_e\Bigr\}.
\label{eq:occ_feasible}
\end{equation}

Whenever a trial vector $\bx$ violates these constraints, SPG first
maps it back to an allowable $\mathsf{n}\in\Omega$ by a projection
step~\cite{Calamai1987}
\begin{equation}
P_w(x_{i\bk}) = \mathrm{clip}\bigl(x_{i\bk} - \lambda\, w_{\bk},\, 0,\, 1\bigr),
\label{eq:Pw}
\end{equation}
where $\mathrm{clip}(y,0,1)$ truncates $y$ to $[0,1]$. 
The scalar $\lambda$ is fixed by bisection on the monotone, piecewise-linear function
\begin{equation}
G(\lambda) \equiv \sum_{i \bk} w_{\bk}\, P_w(x_{i\bk}) - N_e = 0,
\label{eq:Gw}
\end{equation}
so that the projected occupations sum to $N_e$.
The coefficient $\lambda w_{\bk}$ in Eq.~\eqref{eq:Pw} makes $P_w(\bx)$ the
allowable occupation vector closest to $\bx$, in the sense of minimizing
$\sum_{i\bk}(n_{i\bk}-x_{i\bk})^2$ over $\mathsf{n}\in\Omega$.
Appendix~\ref{app:SPG_descent} derives Eq.~\eqref{eq:Pw} and Eq.~\eqref{eq:Gw} and verifies
compatibility with SPG \cite{Birgin2000}.

At $\mathsf{n}_0\in\Omega$, the energy gradient $\bg=\nabla_{\mathsf{n}} E$ has
components $g_{i\bk}=\partial E/\partial n_{i\bk}$.
To decide whether the inner loop has converged, SPG uses the
projected-gradient map of Birgin \textit{et al.}~\cite{Birgin2000}
\begin{equation}
\mathsf{g}_t(\mathsf{n}) \equiv P_w(\mathsf{n} - t\,\bg) - \mathsf{n},
\qquad t>0,
\label{eq:gt_def}
\end{equation}
which measures how much a raw gradient step of size $t$ must be
corrected to restore the constraints.
When $t=1$, a small value of $\|\mathsf{g}_1(\mathsf{n}_0)\|_\infty$ signals
that $\mathsf{n}_0$ is close to a constrained minimum and no further update is
needed, where $\| \cdot \|_\infty$ denotes the infinity norm (or the maximum norm).

\begin{figure}[htbp]
  \centering
  \begin{tikzpicture}[scale=1.15, >=stealth, thick,
    every node/.style={font=\footnotesize}]
    \fill[blue!12] (2.4,1.7) ellipse (2.15 and 1.35);
    \draw[blue!55!black] (2.4,1.7) ellipse (2.15 and 1.35);
    \node[blue!60!black] at (1.4,2.5) {$\Omega$};
    \coordinate (n0) at (3.35,1.35);
    \coordinate (x)  at (5.55,0.55);
    \coordinate (np) at (4.12,2.48);
    \draw[->, red!75!black] (n0) -- (x)
      node[midway, below left, yshift=-1pt]
      {\color{red!75!black}$-\alpha_k\bg$};
    \draw[->, teal!80!black, densely dashed] (x) -- (np)
      node[midway, right, xshift=2pt]
      {\color{teal!80!black}$P_w$};
    \draw[->, blue!70!black, very thick] (n0) -- (np)
      node[midway, left, xshift=-2pt]
      {\color{blue!70!black}$\mathsf{n}(\lambda)$};
    \fill[blue!70!black] (n0) circle (1.7pt)
      node[below left] {$\mathsf{n}_0$};
    \fill[red!75!black] (x) circle (1.7pt)
      node[below right] {$\mathsf{n}_0-\alpha_k\bg$};
    \fill[teal!80!black] (np) circle (1.7pt)
      node[above right] {$\mathsf{n}_0+\bd_k$};
  \end{tikzpicture}
  \caption{Schematic of one SPG occupation step. From
  $\mathsf{n}_0\in\Omega$ near the feasible boundary, a spectral gradient
  step $-\alpha_k\bg$ (red) may leave $\Omega$; the projector $P_w$ (teal)
  returns the trial to $\mathsf{n}_0+\bd_k\in\Omega$. The search then follows
  the feasible chord $\mathsf{n}(\lambda)=\mathsf{n}_0+\lambda\bd_k$ (blue),
  $\lambda\in[0,1]$.}
  \label{fig:SPG_chord}
\end{figure}
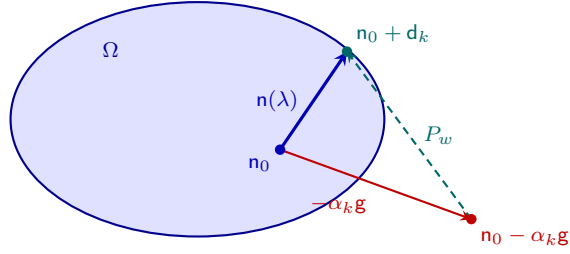

Otherwise one SPG inner iteration proceeds as follows; the integer $k$
labels the iteration ($k=1,2,\ldots$) and is distinct from the Bloch
vector $\bk$.
An adaptive Barzilai-Borwein spectral steplength $\alpha_k$ is taken from the two most recent
accepted iterations~\cite{Barzilai1988,Birgin2000}; on the first step
$\alpha_k=1/\|\mathsf{g}_1(\mathsf{n}_0)\|_\infty$.
The algorithm forms a \emph{feasible chord}: it takes a gradient step, projects
back to $\Omega$, and connects the endpoints by a straight line (see Fig.~\ref{fig:SPG_chord}),
\begin{equation}
\bd_k = \mathsf{g}_{\alpha_k}(\mathsf{n}_0)
= P_w(\mathsf{n}_0 - \alpha_k \bg) - \mathsf{n}_0, \qquad
\mathsf{n}(\lambda) = \mathsf{n}_0 + \lambda\,\bd_k, \quad \lambda\in[0,1].
\label{eq:SPG_chord}
\end{equation}
Because both $\mathsf{n}_0$ and $\mathsf{n}_0+\bd_k$ lie in $\Omega$, every
$\lambda\in[0,1]$ obeys the box and electron-count constraints exactly,
as schematically illustrated in Fig.~\ref{fig:SPG_chord}.
The energy $E(\lambda)\equiv E(\mathsf{n}(\lambda))$ is then reduced along this
chord, starting from $\lambda_0=1$. 

A trial $\lambda$ is accepted when it satisfies the Armijo
sufficient-decrease test~\cite{Armijo1966,Birgin2000,Grippo1986},
\begin{equation}
E(\lambda) \le f_{\max} + \gamma\,\lambda\,\phi'(0), \qquad
\phi'(0) = \bg^\top \bd_k,
\label{eq:SPG_armijo}
\end{equation}
where $\phi'(0)$ is the initial energy derivative along the chord ($\lambda=0$),
$\gamma>0$, and
$f_{\max}=\max\{E(\mathsf{n}^{(k-j)})\mid 0\le j\le \min\{k,M-1\}\}$;
$\mathsf{n}^{(k-j)}$ denotes the occupation vector at inner iteration~$k-j$.
For $M>1$ this is the Grippo-Lampariello-Lucidi nonmonotone rule \cite{Grippo1986}, which
can help the optimizer cross shallow barriers; for $M=1$ only $j=0$ enters
and $f_{\max}=E(\mathsf{n}_0)$, reducing Eq.~\eqref{eq:SPG_armijo} to the usual
monotone Armijo condition~\cite{Armijo1966}. In this work, we have set $M=1$ for the monotone Armijo condition.
If a trial with $\lambda>1$ leaves $\Omega$, it is repaired by
$P_w$. 

\subsubsection{Orbital optimization on generalized Stiefel manifolds}
\label{subsub:orbopt}
For the orbital sub-problem with fixed occupations, the orbital coefficients are constrained by
Eq.~\eqref{eq:orb_constraint}, so Euclidean updates must be replaced by
geometry-aware steps. Following our previous work \cite{Luo2025}, natural orbital coefficients in a given basis live on the 
generalized Stiefel manifold,
\begin{equation}
\mathrm{St}(N_b, N_{\mathrm{basis}}; \mathsf{S}) = \bigl\{ \mathsf{X} \in \mathbb{C}^{N_{\mathrm{basis}} \times N_b} : \mathsf{X}^\dagger \mathsf{S}\mathsf{X} = \mathsf{I}_{N_b} \bigr\},
\end{equation}
with $N_{\mathrm{basis}}$ the number of basis functions per band. 
For planewave calculations with norm-conserving pseudopotentials, the overlap matrix is the identity, $\mathsf{S} = \mathsf{I}_{N_{\mathrm{basis}}}$.
More generally, for $N_\mathrm{ao}$ atomic orbital (AO) basis functions, by applying a Cholesky factorization of $\mathsf{S}$,
the generalized Stiefel manifold $\mathrm{St}(N_b, N_{\mathrm{ao}}; \mathsf{S})$ can be mapped
to the standard Stiefel manifold.
At $\mathsf{X} \in \mathrm{St}(N_b, N_{\mathrm{basis}}; \mathsf{I}_{N_{\mathrm{basis}}})$, we have the tangent space
\begin{equation}
    T_{\mathsf{X}} \mathrm{St} = \left\{ \mathsf{Y} = \mathsf{X} \mathsf{B} + \mathsf{Z} \mid \mathsf{B}^{\dagger} = -\mathsf{B},\ \mathsf{Z}^{\dagger} \mathsf{X} = 0 \right \},
    \label{eq:tangent_manifold}
\end{equation}
where $\mathsf{Y}, \mathsf{Z} \in \mathbb{C}^{N_{\mathrm{basis}} \times N_b}$, $\mathsf{B} \in \mathbb{C}^{N_b\times N_b}$.
Here,
$\mathsf{B}$ is a skew-Hermitian matrix and $\mathsf{Z}$ is a matrix orthogonal to $\mathsf{X}$.
For the Stiefel manifold, the canonical metric is used
\begin{equation}
    \langle \mathsf{U}, \mathsf{V}\rangle_{\mathsf{X}}^{c} = \tr \left[ \mathsf{U}^{\dagger} \left( \mathsf{I}_{N_{\mathrm{basis}}} - \half \mathsf{X} \mathsf{X} ^{\dagger} \right) \mathsf{V}\right]
    \label{eq:canonical_metric}
\end{equation}
where $\mathsf{U}, \mathsf{V}\in T_{\mathsf{X}} \mathrm{St}$.
The orthogonal projection of any vector $\mathsf{V}\in \mathbb{C}^{N_\mathrm{basis} \times N_b}$ onto
 $T_{\mathsf{X}} \mathrm{St}$ is
\begin{equation}
    \pi_{\mathsf{X}} (\mathsf{V}) = \mathsf{V}- \mathsf{X} \sym (\mathsf{X} ^{\dagger} \mathsf{V}) \quad \text{with} \quad \sym(A) = \tfrac{1}{2}(A + A^\dagger).
    \label{eq:projection}
\end{equation}

Two essential components of retraction and vector transport are needed in the Riemannian optimization (see details in \cite{Luo2025}).
In this work, we used the polar decomposition as the retraction and the vector transport by projection as the vector transport.
For multiple-k points, the coefficients are stored using the product of Stiefel manifolds as done in \cite{Luo2025},
\begin{equation}
\prod_{\bk=1}^{N_{\mathrm{k}}} \mathrm{St}(N_b, N_{\mathrm{basis}}; \mathsf{I}_{N_{\mathrm{basis}}}).
\label{eq:prodmanifold}
\end{equation}

The Euclidean gradient in the ambient coefficient space is $\nabla E$ (its argument of $\mathsf{C}_{\bk}$ is omitted for simplicity)
the corresponding Riemannian gradient $\mathsf{g} = \grad E$ in the Stiefel manifold, can be computed with the Euclidean gradient 
\begin{equation}
    \grad E = \nabla E - \mathsf{X} \left( \nabla E \right)^\dagger \mathsf{X}.
    \label{eq:Riemannian_gradient}
\end{equation}
In the conjugate gradient method, the new search direction $\mathsf{d}_{k+1}$ is computed as 
\begin{equation}
    \mathsf{d}_{k+1} = - \mathsf{g}_{k+1} + \beta_{k+1}    \mathcal{T}_{\alpha_k \mathsf{d}_{k}} (\mathsf{d}_k),
    \label{eq:conjugate_gradient}
\end{equation}
where $\mathcal{T}_{\alpha_k \mathsf{d}_{k}} (\mathsf{d}_k)$ is the projection based vector transport. Again,
we used the Polak-Ribi\`ere form for the $\beta$ parameter in the conjugate gradient method.

\begin{figure}
  \centering
  \includegraphics[width=0.6\textwidth]{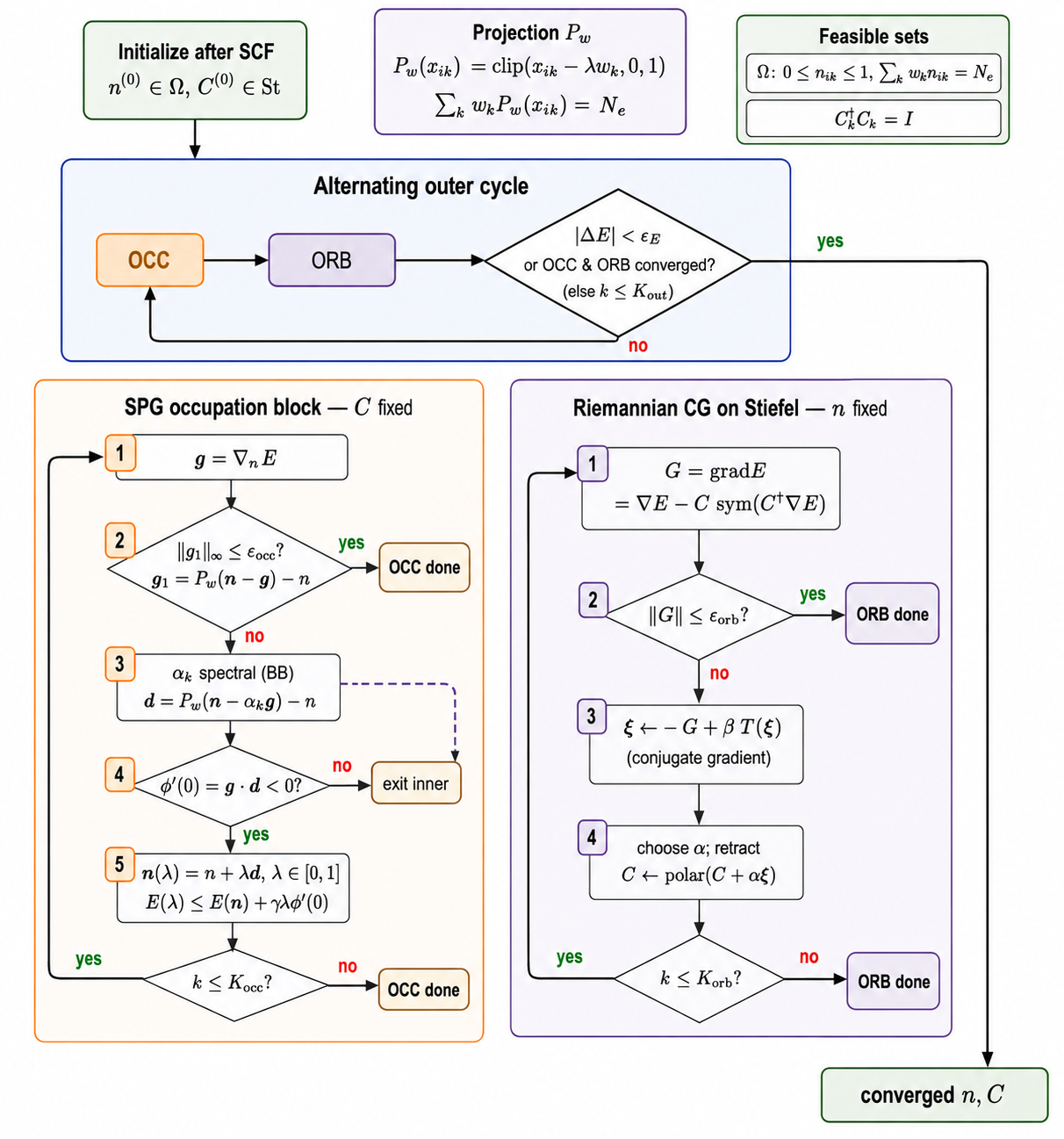}
  \caption{Schematic illustration of the RDMFT algorithm with the SPG occupation 
  optimizer and the Riemannian conjugate gradient orbital optimizer.}
  \label{fig:flowchart}
\end{figure}

Combining these two optimization strategies, our RDMFT algorithm can be illustrated in Fig.~\ref{fig:flowchart}. 
Such algorithmic design allows easy substitution of alternative methods, e.g. in the occupation part the SPG method can be 
replaced with the EBI method. The outer maximum iteration $K_{\mathrm{out}}$ and the inner maximum iterations $K_{\mathrm{occ}}$
and $K_{\mathrm{orb}}$ are used to control the number of alternating outer cycles and the effort spent in each occupation and orbital subproblem, respectively.
Besides the energy tolerance $\epsilon_\mathrm{E}$ for the outer loop,
the gradient tolerances $\epsilon_{\mathrm{orb}}$ and $\epsilon_{\mathrm{occ}}$
set the convergence thresholds for the orbital and occupation subproblems, respectively.
Post-processing of the natural occupations and the natural orbitals can be done after the convergence has been reached;
further observables are discussed in Sec.~\ref{sec:summary}.

\section{Implementation and Computational details}
\label{sec:computational_details}

The algorithms of Sec.~\ref{sub:minimization}, with occupation updates in
Sec.~\ref{subsub:occ_opt} and orbital updates in Sec.~\ref{subsub:orbopt}, are
realized in the
planewave QE package version 7.5  as a self-consistent-field
variational minimization\cite{Giannozzi2009}. 
After a converged Kohn-Sham or hybrid calculation,
the solver updates the natural occupations $n_{i\bk}\in[0,1]$ and the natural
orbitals $\psi_{i\bk}(\br)$ already stored as planewave coefficient
matrices. 
The energy is iteratively minimized subject to Eq.~\eqref{eq:occ_constraint} and the orthonormality constraint of
Eq.~\eqref{eq:orb_constraint}. The initial occupation numbers are seeded from a 
KS calculation using the smearing method to start with Fermi-Dirac initial guess.
The initial orbital coefficients can be seeded from a few-step KS calculation such that initial orbitals 
are orthonormalized. Given the minimal cost of KS calculation, starting with a converged set of KS orbitals is 
recommended as the kinetic energy has been largely suppressed.

To make sure that the implementation is correct, the analytical derivatives with respect to the occupation numbers 
and the orbital coefficients are calculated and compared with the finite difference method to a very high accuracy.
The consistency between the objective function and its gradients is also crucial for the line search to work properly.

All the calculations are performed using the modified planewave QE package as noted above. Natural orbitals 
are all expanded in the planewave basis with an energy cutoff of 70 Ry. 
The norm-conserving pseudopotentials  (ONCV) \cite{Schlipf2015} are used to describe the electron-ion interactions.
For molecules and other isolated systems, only the $\Gamma$ point is used in a cubic supercell of length 15.0~\AA.
For charged systems, the spin-polarized calculation is needed and the Martyna-Tuckerman correction is applied \cite{Martyna1999}.
For power-functionals, the exponent is set to $\alpha=0.65$.
Benchmark solid calculations only uses Monkhorst-Pack k-point meshes of $2\times 2\times 2$ unless stated otherwise.
For the Fock operator sampling, we use $2\times 2\times 2$ q-grid. The Coulomb potential divergences at small q vectors
are treated using the Gygi-Baldereschi correction \cite{Gygi1986}. The ACE
algorithm is used in the construction of the EXX. 
All the calculations stop when the energy difference between two consecutive iterations is less than $10^{-8}$ Ry.
Unless stated otherwise, Sec.~\ref{sec:results} uses the settings above with the
following additions: HF and power functional benchmarks on \ce{H2},
\ce{Si}, and \ce{Na} employ a $2\times 2\times 2$ Monkhorst-Pack mesh with offset
$(0.215, 0.35, 0.625)$.
Occupations and orbitals are optimized alternately except for
the test of initial occupation where orbitals are frozen at converged HF values.
RDMFT runs are capped at 50 outer iterations where noted and by default maximum 10 inner iterations is set for both the orbital and the occupation loop.

\section{Results and Discussion}
\label{sec:results}

\subsection{Implementation validation}
\label{sub:implementation_validation}

We first verify the RDMFT implementation against HF references
from converged Kohn-Sham hybrid calculations on three representative systems:
an isolated \ce{H2} molecule, gapped silicon bulk (cubic diamond), and metallic sodium bulk (body-centered cubic).
Computational parameters follow Sec.~\ref{sec:computational_details}; the \ce{H2}
equilibrium bond length is 0.7414~\AA, \ce{Si} uses a two-atom primitive cell with
conventional cubic lattice constant $a=10.236$~a.u.\ ($a/2=5.118$~a.u.), and \ce{Na}
a one-atom primitive cell with cubic lattice constant $a=8.108$~a.u.

For RDMFT, we compare the SPG and EBI occupation optimizers with the same Riemannian conjugate-gradient orbital
update.
To depart from the idempotent KS starting point, initial occupations are
obtained by Gaussian smearing with width 0.1~Ry.
Total energies are listed in Table~\ref{tab:hf_energies} and occupations in
Table~\ref{tab:hf-occupations}.

For \ce{H2}, SPG reproduces the HF energy and occupations exactly
($\Delta E=0$).
EBI agrees to within $0.12\,\mu$Ry while preserving the occupied occupations.
For silicon, SPG again matches the KS reference exactly ($\Delta E=0$).
EBI stops short of full convergence ($\Delta E=+0.71\,\mathrm{mRy}$) and
slightly undershoots valence occupations (e.g.\ $0.9997$ instead of $1.0$ on
several bands) while populating conduction states at the $10^{-4}$ level.
For sodium, SPG remains accurate to $0.01\,\mu$Ry and reproduces the KS
occupations exactly, including the $0.5$ occupation at the Fermi level.
EBI gives $\Delta E=+0.11\,\mathrm{mRy}$ and shifts the Fermi-level occupation
to $0.4998$ while introducing a small occupation ($2.4\times 10^{-4}$) on an
adjacent band.

Overall, SPG is the more reliable choice for these HF benchmarks across
molecules, gapped solids, and metals.
When occupations approach $0$ or $1$, the erf parameterization used in EBI
(Sec.~\ref{subsub:occ_opt}) suppresses occupation updates through vanishing
derivatives, whereas SPG projects directly onto the feasible set and avoids this
issue. Occupation parameterizations can also create artificial saddle points
when occupations are degenerate~\cite{Cartier2025}, typically requiring a
second-order treatment to escape them. Because SPG does not introduce an
auxiliary parameterization, it is not subject to this mechanism; across the
systems tested here we observe no such trapping, though a broader assessment
remains warranted. 

\begin{table}[htbp]
    \centering
    \caption{Total energies from Kohn-Sham hybrid SCF and
    RDMFT with the HF functional. Values in parentheses give
    $\Delta E = E_{\mathrm{RDMFT}} - E_{\mathrm{KS}}$ in $\mu$Ry
    or mRy, as indicated.
    $^\dagger$ marks EBI runs that did not reach the $10^{-8}$~Ry
    convergence threshold.}
    \label{tab:hf_energies}
    \setlength{\tabcolsep}{6pt}
    \begin{tabular}{lccc}
        \hline\hline
        System & KS  & SPG & EBI \\
        \hline
        \ce{H2} &
          \makebox[1.8cm][r]{$-2.26575897$} &
          \makebox[1.8cm][r]{$-2.26575897$}\,{\scriptsize$(+0\,\mu\mathrm{Ry})$} &
          \makebox[1.8cm][r]{$-2.26575885$}\,{\scriptsize$(+0.12\,\mu\mathrm{Ry})$} \\
        \ce{Si} &
          \makebox[1.8cm][r]{$-15.07589955$} &
          \makebox[1.8cm][r]{$-15.07589955$}\,{\scriptsize$(+0\,\mu\mathrm{Ry})$} &
          \makebox[1.8cm][r]{$-15.07518854^{\dagger}$}\,{\scriptsize$(+0.71\,\mathrm{mRy})$} \\
        \ce{Na} &
          \makebox[1.8cm][r]{$-84.49001110$} &
          \makebox[1.8cm][r]{$-84.49001109$}\,{\scriptsize$(+0.01\,\mu\mathrm{Ry})$} &
          \makebox[1.8cm][r]{$-84.48990456^{\dagger}$}\,{\scriptsize$(+0.11\,\mathrm{mRy})$} \\
        \hline\hline
    \end{tabular}
\end{table}

\begin{table}[htbp]
    \centering
    \caption{HF occupations for \ce{H2}, \ce{Si}, and \ce{Na}.
    $i_k$ and $i_b$ are the $k$-point and band indices, respectively.
    EBI deviations concentrate near the Fermi level for \ce{Si} and \ce{Na}.}
    \label{tab:hf-occupations}
    \begin{tabular}{cc *{9}{r}}
    \hline\hline
    \multirow{2}{*}{$i_k$} & \multirow{2}{*}{$i_b$}
     & \multicolumn{3}{c}{\ce{H2}} & \multicolumn{3}{c}{\ce{Si}} & \multicolumn{3}{c}{\ce{Na}} \\
    \cline{3-5} \cline{6-8} \cline{9-11}
     &  & KS & SPG & EBI & KS & SPG & EBI & KS & SPG & EBI \\
    \hline
    1 & 1 & 1.000000 & 1.000000 & 1.000000 & 1.000000 & 1.000000 & 1.000000 & 1.000000 & 1.000000 & 1.000000 \\
    1 & 2 & 0.000000 & 0.000000 & 0.000000 & 1.000000 & 1.000000 & 0.999665 & 1.000000 & 1.000000 & 1.000000 \\
    1 & 3 & 0.000000 & 0.000000 & 0.000000 & 1.000000 & 1.000000 & 0.999665 & 1.000000 & 1.000000 & 1.000000 \\
    1 & 4 & 0.000000 & 0.000000 & 0.000000 & 1.000000 & 1.000000 & 0.999665 & 1.000000 & 1.000000 & 1.000000 \\
    1 & 5 &  & 0.000000 & 0.000000 & 0.000000 & 0.000000 & 0.000144 & 1.000000 & 1.000000 & 0.999778 \\
    1 & 6 &  &  &  & 0.000000 & 0.000000 & 0.000144 & 0.000000 & 0.000000 & 0.000000 \\
    1 & 7 &  &  &  & 0.000000 & 0.000000 & 0.000144 & 0.000000 & 0.000000 & 0.000000 \\
    1 & 8 &  &  &  & 0.000000 & 0.000000 & 0.000068 & 0.000000 & 0.000000 & 0.000000 \\
    1 & 9 &  &  &  &  &  &  & 0.000000 & 0.000000 & 0.000000 \\
    2 & 1 &  &  &  & 1.000000 & 1.000000 & 1.000000 & 1.000000 & 1.000000 & 1.000000 \\
    2 & 2 &  &  &  & 1.000000 & 1.000000 & 1.000000 & 1.000000 & 1.000000 & 1.000000 \\
    2 & 3 &  &  &  & 1.000000 & 1.000000 & 0.999769 & 1.000000 & 1.000000 & 1.000000 \\
    2 & 4 &  &  &  & 1.000000 & 1.000000 & 0.999769 & 1.000000 & 1.000000 & 1.000000 \\
    2 & 5 &  &  &  & 0.000000 & 0.000000 & 0.000193 & 0.500000 & 0.500000 & 0.499794 \\
    2 & 6 &  &  &  & 0.000000 & 0.000000 & 0.000087 & 0.000000 & 0.000000 & 0.000243 \\
    2 & 7 &  &  &  & 0.000000 & 0.000000 & 0.000087 & 0.000000 & 0.000000 & 0.000000 \\
    2 & 8 &  &  &  & 0.000000 & 0.000000 & 0.000000 & 0.000000 & 0.000000 & 0.000000 \\
    2 & 9 &  &  &  &  &  &  & 0.000000 & 0.000000 & 0.000000 \\
    3 & 1 &  &  &  & 1.000000 & 1.000000 & 1.000000 & 1.000000 & 1.000000 & 1.000000 \\
    3 & 2 &  &  &  & 1.000000 & 1.000000 & 1.000000 & 1.000000 & 1.000000 & 1.000000 \\
    3 & 3 &  &  &  & 1.000000 & 1.000000 & 0.999868 & 1.000000 & 1.000000 & 1.000000 \\
    3 & 4 &  &  &  & 1.000000 & 1.000000 & 0.999868 & 1.000000 & 1.000000 & 1.000000 \\
    3 & 5 &  &  &  & 0.000000 & 0.000000 & 0.000279 & 0.000000 & 0.000000 & 0.000000 \\
    3 & 6 &  &  &  & 0.000000 & 0.000000 & 0.000279 & 0.000000 & 0.000000 & 0.000000 \\
    3 & 7 &  &  &  & 0.000000 & 0.000000 & 0.000000 & 0.000000 & 0.000000 & 0.000000 \\
    3 & 8 &  &  &  & 0.000000 & 0.000000 & 0.000000 & 0.000000 & 0.000000 & 0.000000 \\
    3 & 9 &  &  &  &  &  &  & 0.000000 & 0.000000 & 0.000000 \\
    \hline\hline
    \end{tabular}
\end{table}

\begin{figure}[h]
    \centering
    \includegraphics[width=0.8\textwidth]{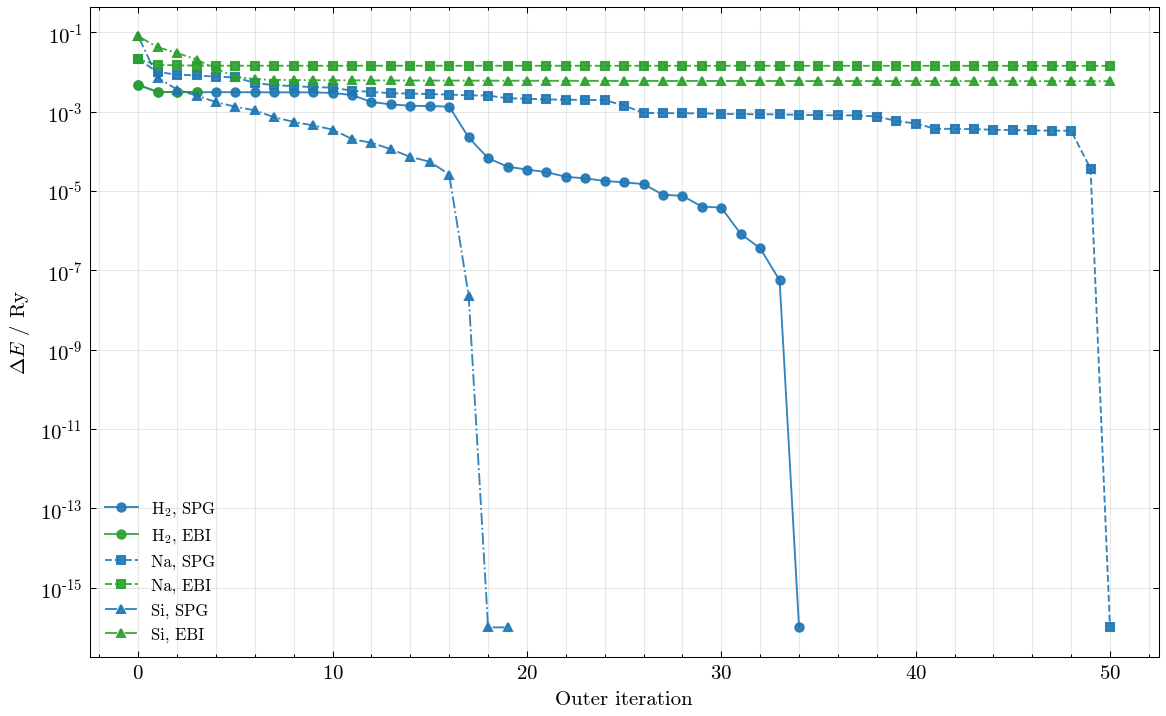}
    \caption{Energy convergence for the power functional with SPG and EBI
    occupation optimizers.
    The reference energy is the lower value between the two optimizers at each
    system.}
    \label{fig:energy_convergence_power}
\end{figure}

We next compare SPG and EBI with the power functional on
the same three systems.
Table~\ref{tab:power_energies} lists the RDMFT total energies after up to 50
outer iterations.
SPG yields lower energies than EBI for all three systems.
After 50 outer iterations, EBI has not reached the $10^{-8}$~Ry threshold for \ce{Si}
and \ce{Na}; SPG also remains above threshold for \ce{Na} but is already lower in energy
than EBI.
Figure~\ref{fig:energy_convergence_power} shows the energy as a function of
outer iteration, using the lower of the two optimizer energies as reference.
SPG converges monotonically to the lower values, whereas EBI stagnates at
higher energy and is unlikely to converge to the lower plateau even with larger outer iterations.

\begin{table}[h]
    \centering
    \caption{RDMFT total energies with the power functional.
    $^\dagger$ denotes runs that reached 50 outer iterations without meeting
    the $10^{-8}$~Ry convergence criterion.}
    \label{tab:power_energies}
    \setlength{\tabcolsep}{6pt}
    \begin{tabular}{lcc}
        \hline\hline
        System & SPG & EBI \\
        \hline
        \ce{H2} &
          \makebox[1.8cm][r]{$-2.26891386$} &
          \makebox[1.8cm][r]{$-2.26575948$} \\
        \ce{Si} &
          \makebox[1.8cm][r]{$-15.19544633$} &
          \makebox[1.8cm][r]{$-15.18957380^{\dagger}$} \\
        \ce{Na} &
          \makebox[1.8cm][r]{$-84.61019266^{\dagger}$} &
          \makebox[1.8cm][r]{$-84.59587115^{\dagger}$} \\
        \hline\hline
    \end{tabular}
\end{table}

To test sensitivity to the starting occupations, we apply a controlled
perturbation $\epsilon$ to bands near the Fermi level: occupations below $0.5$
are increased by $\epsilon$ and those above $0.5$ are decreased by $\epsilon$.
Initial orbitals are taken from a converged HF calculation and held fixed to
isolate occupation optimization.
Using the power functional on bulk \ce{Na}, we vary $\epsilon$ from $10^{-4}$ to
$10^{-1}$ for both SPG and EBI. For occupation optimization only, SPG returns the same energy within $\sim$70 inner
iterations for all perturbation strengths, whereas EBI yields distinct,
non-converged energies after 500 inner iterations.

\subsection{Molecules}
\label{sub:molecules}
The same RDMFT formulation developed above applies without change to isolated
molecules and to periodic solids.
We compare fractionally charged \ce{LiH} with published RDMFT data~\cite{Sharma2008}.
Here the total electron number is varied through a charge parameter $\delta$
(equivalently $N_e = 4 + \delta$ for \ce{LiH}).
Quantitative agreement with the hybrid KS reference for this charged molecule
requires a spin-polarized treatment together with the Martyna--Tuckerman
correction for electrostatic finite-size effects~\cite{Martyna1999}
(Sec.~\ref{sec:computational_details}); the planewave cutoff is 70~Ry.
RDMFT is then iterated until the energy criterion of
Sec.~\ref{sec:computational_details} is met.

\begin{figure}[htbp]
    \centering
    \includegraphics[width=0.8\textwidth]{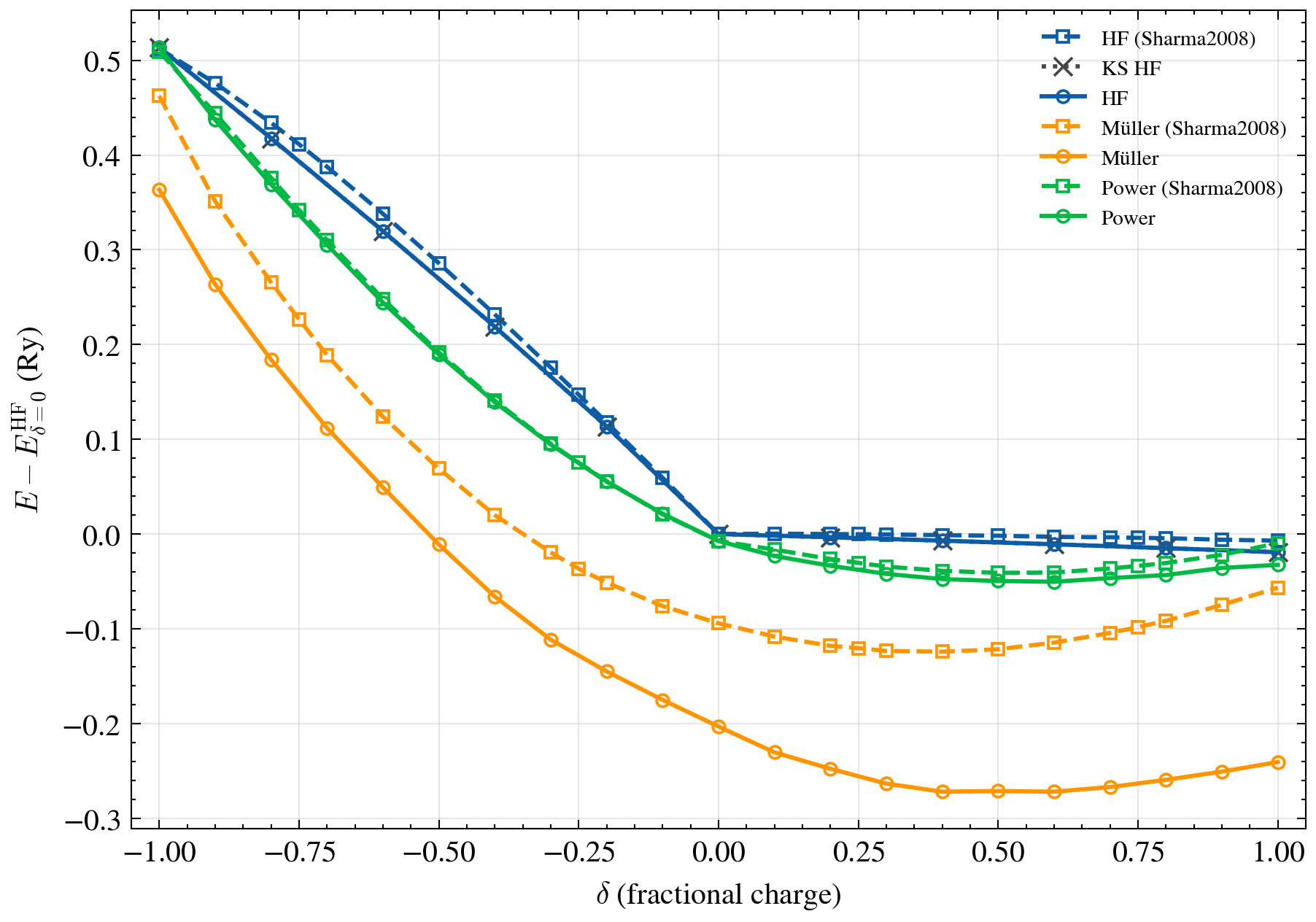}
    \caption{Fractionally charged \ce{LiH}: RDMFT total energies relative to
    $E_{\delta=0}^{\mathrm{HF}}$.
    Black: HF reference using KS hybrid loop; blue: present HF (SPG); green: power
    functional (SPG); orange: M\"uller functional (SPG).
    Dashed curves: digitized data from Sharma \textit{et al.}~\cite{Sharma2008}.}
    \label{fig:energy_convergence_lih}
\end{figure}

As a reference, we first reproduce the standard HF curve.
Published results of Sharma \textit{et al.}~\cite{Sharma2008} are digitized for
comparison.
Because all-electron and pseudopotential calculations differ in absolute
energy, all curves are shifted by subtracting the neutral HF energy
$E_{\delta=0}^{\mathrm{HF}}$.
Figure~\ref{fig:energy_convergence_lih} shows that our HF results (SPG
occupation optimizer) agree with the KS reference to within numerical precision
and display the expected piecewise concave shape~\cite{Mori-Sanchez2008,Li2017}.
The digitized Sharma data lie slightly above this reference, which is indicative
of incomplete convergence.

With the power functional, our RDMFT energies largely agree with Sharma
\textit{et al.}~\cite{Sharma2008} and are slightly lower at several charge
points.
For the M\"uller functional, our converged curve lies substantially below the
published data; the largest difference at $N_e=5.0$ ($\delta=1.0$) is about
0.1~Ha.
M\"uller converges more slowly than the power functional in these runs, and the
huge discrepancy is again likely due to their failure in achieving convergence.
Further investigation is needed to figure out the exact reason and is left for future.
These comparisons show that the present SPG-based algorithm converges more
reliably and to lower energies than the earlier solid-state RDMFT data for
fractionally charged \ce{LiH}, and thereby provides a more trustworthy platform
for studying the derivative discontinuity and for probing convexity or
concavity of the underlying XC functional.

\begin{figure}[htbp]
  \centering
  \includegraphics[width=0.9\textwidth]{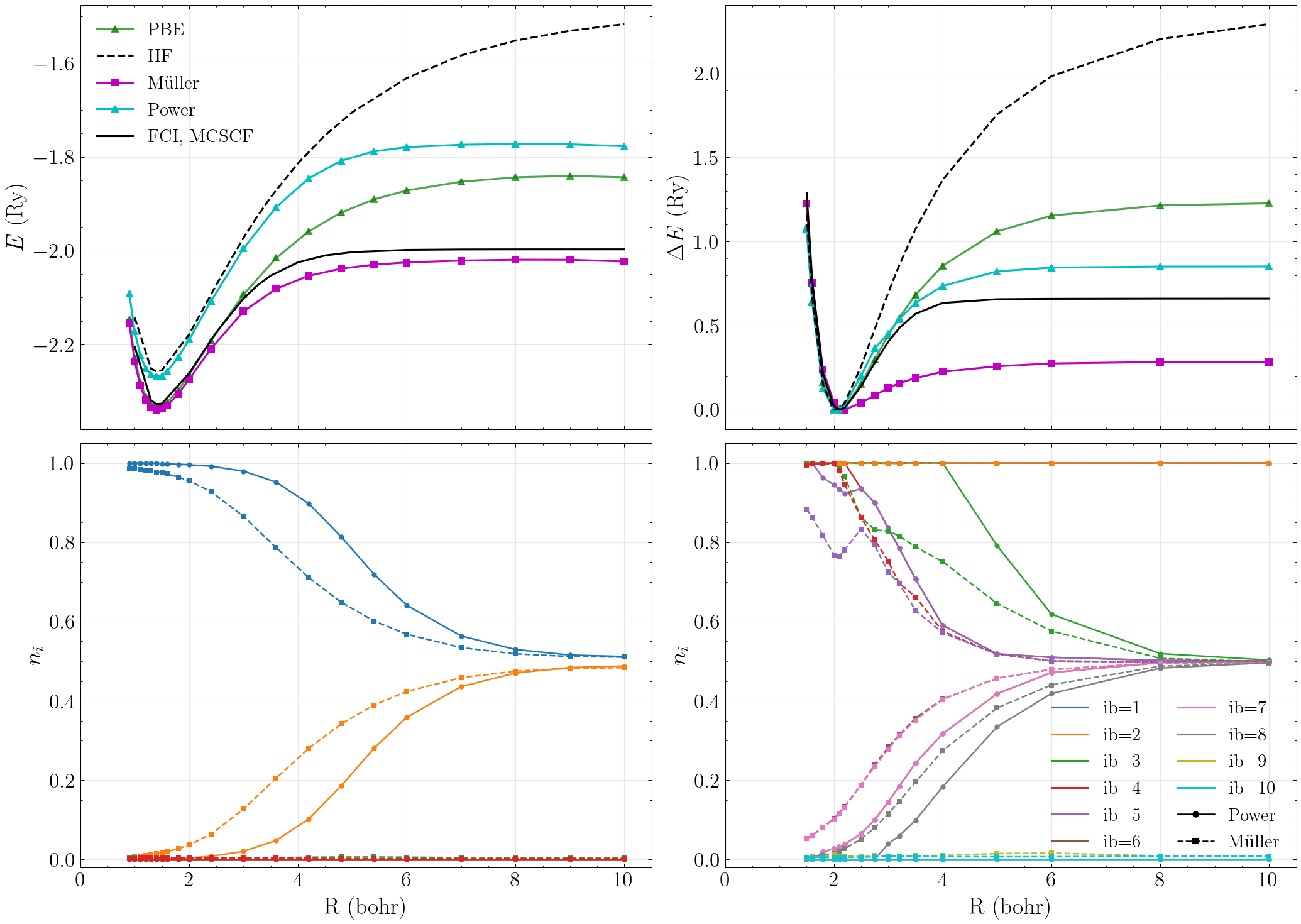}
  \caption{Dissociation energy curves for \ce{H2} and \ce{N2} molecules.
  Full-CI for \ce{H2} and MCSCF for \ce{N2} are used as references. Dashed black curves are the RHF results \cite{Wang2022}.
  Power functional and M\"uller functional in RDMFT are presented with the KS PBE results. For \ce{N2}, the global minimum of each curve is subtracted so that these curves are on the same scale.
  The natural occupation numbers are shown in the bottom panels for different bond lengths in our RDMFT calculations.
  }
  \label{fig:dissociation_summary}
\end{figure}

Bond dissociation provides a stringent test of RDMFT because the correct
description requires fractional natural occupations on the dissociation limit.
Figure~\ref{fig:dissociation_summary} compares the total energy curves  of PBE \cite{PBE1996} in DFT, power and
M\"uller functional in RDMFT for stretched
\ce{H2} and \ce{N2} against quantum-chemistry references \cite{Wang2022}, full configuration interaction (full CI) for \ce{H2} and multi-configurational self-consistent field (MCSCF) for \ce{N2}.
All calculations are spin-unpolarized. For \ce{H2} molecule, the M\"uller functional yields close dissociation curve compared to HF,  PBE,  and power functional. However, for the \ce{N2} molelcule, M\"uller gives much lower energies in the dissociation limit, which is consistent with earlier studies \cite{Gritsenko2005}. 
Power functional gives a correct shape but its performance depends on the  adopted parameter $\alpha$ for each system. For \ce{H2}, there are two major natural orbitals whose occupation numbers are far away from boundaries (one strongly occupied and one weakly occupied).  RDMFT give both half occupations in the dissociation limit, and  the occupation, of either strongly occupied or weakly occupied orbital, 
of M\"uller is clearly closer to the half-occupancy than that of power at all bond lengths, which can be regarded as an indicator of stronger correlation. For \ce{N2}, five such orbitals are involved, their occupations all approach the half occupancy. Again, the occupation of the same band exhibits similar trend as seen in \ce{H2}.

\subsection{Solids}
\label{sub:solids}
\begin{figure}[!htbp]
    \centering
    \includegraphics[width=0.8\textwidth]{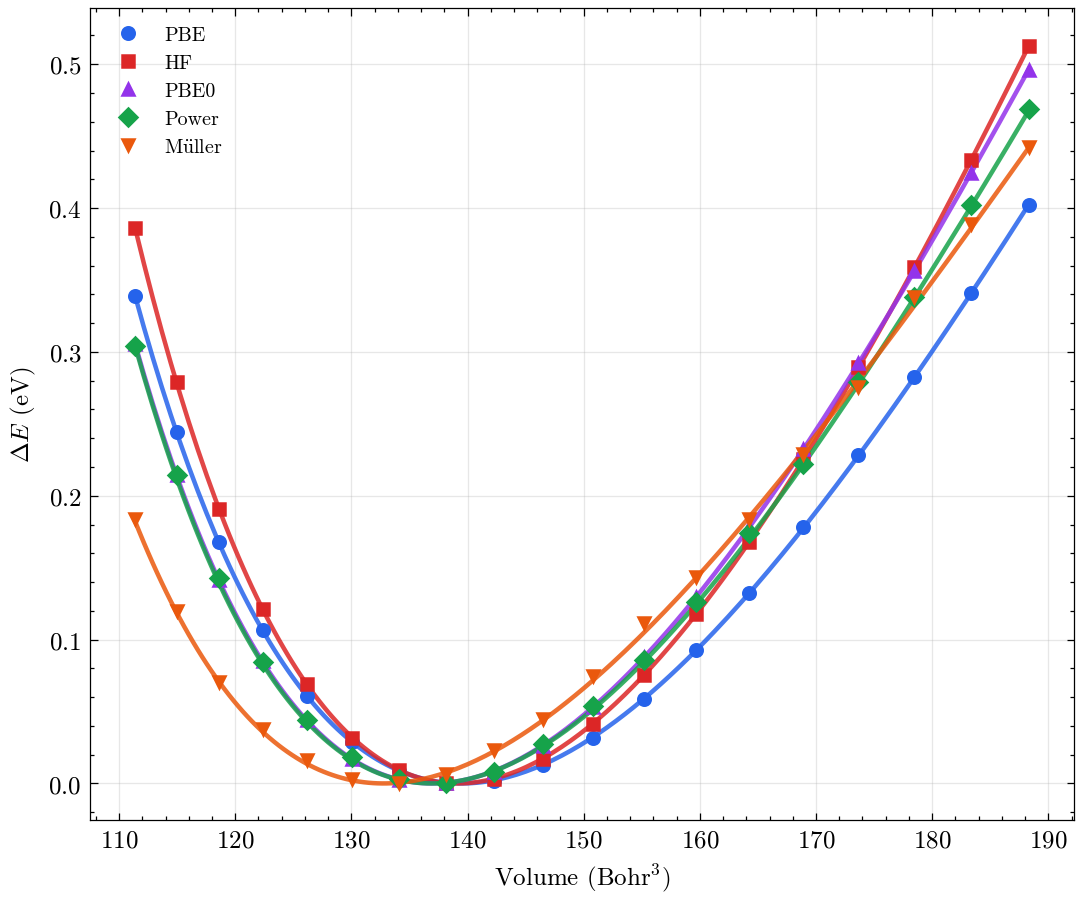}
    \caption{Silicon EOS from RDMFT power and M\"uller functionals,
     compared with PBE, HF, and PBE0 references.
    Energy and volume are per atom. Energy is relative to the equilibrium energy per atom.}
    \label{fig:equation_of_states_si}
  \end{figure}

We next examine bulk silicon in the diamond structure as a representative
solid. To be more realistic, the silicon equation-of-state scans use
$4\times 4\times 4$ with offset $(0.25, 0.25, 0.25)$.
Table~\ref{tab:si-eos-bm3} summarizes fits from PBE, HF, and PBE0
self-consistent calculations together with RDMFT results for the power and
M\"uller functionals and experimental reference data; Fig.~\ref{fig:equation_of_states_si} plots the raw
energy-volume data.
Among the Kohn-Sham references, PBE and HF slightly overestimate the
equilibrium lattice constant ($a_0 = 10.37$ and $10.36$~a.u., respectively,
vs.\ $10.26$~a.u. experimentally), whereas PBE0 is closest to experiment
($10.31$~a.u.).
The present RDMFT power-functional fit gives the same $a_0 = 10.31$~a.u. as
PBE0, with $B_0 = 94.5$~GPa and $B_0' = 4.54$, placing the bulk modulus just
below the experimental value ($97.9$~GPa)~\cite{McSkimin1964} and between PBE
($86.9$~GPa) and HF ($105.2$~GPa).
By contrast, the equilibrium $a_0$ reported by Sharma \textit{et al.}~\cite{Sharma2008}
for the power functional with $\alpha = 0.70$ ($10.55$~a.u.) is expanded by
roughly 2.8\% relative to experiment.
The present M\"uller fit yields a slightly contracted lattice
($a_0 = 10.20$~a.u.) and a substantially softer $B_0 = 81.2$~GPa with
$B_0' = 4.84$, whereas the Sharma \textit{et al.}\ value ($10.49$~a.u.) lies on
the expanded side of experiment.
These trends are visible in Fig.~\ref{fig:equation_of_states_si}, where the
M\"uller curve is shallower near equilibrium than HF or power. Given  the huge discrepancy for the 
M\"uller curve of \ce{LiH} molecule, we meticulously converge the calculation for this solid system as we 
have done for the molecular system. It is likely that the overestimated $10.49$~a.u. is an artifact of 
unconverged calculation instead of the trait of the functional itself. Also M\"uller functional,  
well-known for its overcorrelation ought to yield a contracted lattice constant.
This is clearly seen in our converged M\"uller curve with $a_0 = 10.20$~a.u.

Lastly, we have also attempted  to calculate a typical transition metal oxide, \ce{NiO}  in the power functional. 
Our preliminary results show agreement in local moments of 1.38 $\mu_B$ versus reported 1.36 $\mu_B$
in the antiferromagnetic ordering \cite{Sharma2013}. Further investigation of the spectral 
properties is underway.

\begin{table}[htbp]
  \centering
  \caption{Third-order Birch-Murnaghan EOS fit parameters for diamond \ce{Si}.
  $a_0$ is the conventional eight-atom cubic lattice constant in a.u.;
  $E_0$ is the equilibrium energy per atom (eV/atom).
  Experimental $B_0$ in GPa and $B_0'$ from ultrasonic data at ambient pressure~\cite{McSkimin1964}.
  PBE, HF, and PBE0 results are from self-consistent Kohn-Sham calculations;
  unmarked Power and M\"uller rows are from the present RDMFT calculations.}
  \label{tab:si-eos-bm3}
  \begin{tabular}{lcccc}
    \hline\hline
    Method &
    $a_0$  &
    $B_0$ &
    $B_0'$ &
    $E_0$  \\
    \hline
    Expt.  & 10.26 & 97.9 & 4.24 &  \\
    PBE    & 10.37 & 86.9 & 4.34 & $-107.165$ \\
    HF     & 10.36 & 105.2 & 3.91 & $-103.542$ \\
    PBE0   & 10.31 & 96.9 & 4.19 & $-107.237$ \\
    Power($\alpha=0.70$)$^{\mathrm{a}}$ & 10.55 &  &  &  \\
    Power  & 10.31 & 94.5 & 4.54 & $-104.395$ \\
    M\"uller$^{\mathrm{a}}$ & 10.49 &  &  &  \\
    M\"uller & 10.20 & 81.2 & 4.84 & $-107.840$ \\
    \hline\hline
    \multicolumn{5}{@{}l@{}}{\footnotesize $^{\mathrm{a}}$ Ref.~\cite{Sharma2008}.}
  \end{tabular}
\end{table}

\section{Summary and outlook}
\label{sec:summary}

We have formulated RDMFT for periodic solids in a basis-independent way and
implemented it as a planewave post-SCF module in QE that reuses
ACE for EXX{} handling.
The solver follows an alternating occupation-orbital strategy with a modular
design: occupation and orbital subproblems can be swapped independently.
For occupations we adapt spectral projected gradient (SPG) to periodic systems,
with Euclidean projection onto the weighted feasible set $\Omega$, and also
support first-order EBI with an erf parameterization; for orbitals we use
Riemannian conjugate-gradient updates on generalized Stiefel manifolds that
enforce natural-orbital orthonormality.
Numerical tests show that SPG reliably reproduces HF references on
molecules and solids, converges lower with the power functional, and remains
insensitive to deliberate perturbations of the initial occupations, whereas EBI
can stall when occupations approach $0$ or $1$. 
Fractionally charged \ce{LiH} and the silicon equation of state demonstrate
applications beyond idempotent HF, including comparison with earlier
RDMFT data and with standard DFT references.
Our SPG occupation optimizer is more robust than the first-order EBI for 
both periodic and isolated systems. To further assess the 
original more-expensive second-order EBI in periodic systems and  the box-aware gradient descent 
of \textsc{elk},  a more systematic comparison study of these occupation optimizers 
is left for future work.

The natural next steps toward more efficient calculations include better
preconditioners for the occupation and orbital blocks, fewer ACE rebuilds for
large $k$-meshes, and support for a wider class of multi-channel XC kernels.
A stable optimizer also matters for functional development: when the solver
converges reliably, one can compare XC approximations without wondering whether
an apparent failure is only an incomplete minimization.
The present implementation should make such comparisons easier for solids,
and we hope it will help in refining existing RDMFT functionals and in
constructing new ones that better describe strongly correlated systems that
cannot be described by standard DFT.


Beyond total energies, the present framework naturally provides access to several
experimentally relevant observables. The electron momentum density follows 
directly from the 1-RDM and can be compared with Compton scattering and $(e,2e)$ 
experiments. Extending the method to describe spectral properties \cite{Sharma2013}, such as the 
density of states and spectral functions, remains an important challenge,
requiring either a dynamical formulation or a suitable reconstruction from 
the optimized natural orbitals and occupation numbers. The current 
occupation-orbital optimization framework is also readily applicable to 
spin-polarized and magnetic systems, including open-shell molecules and magnetic 
solids. Finally, incorporating atomic forces and linear-response theory would 
enable molecular dynamics as well as the calculation of phonons, dielectric 
properties, and other response functions within periodic RDMFT.

\section*{Acknowledgments}
K. Luo is grateful for the auspices of the university internal grant No. AS392201.
X.~G. Ren acknowledges the funding support from Research and Development Program of China  (Grant
Nos. 2023YFA1507004 and 2022YFA1403800), and from the National Natural Science Foundation of
China (Grant Nos. 12374067, 12134012, and 12188101).


\appendix

\section{Projector $P_w$ in SPG}
\label{app:SPG_descent}
SPG~\cite{Birgin2000} uses the Euclidean inner product throughout the inner
loop: the chord slope is $\phi'(0)=\bg^\top\bd_k$ Eq.~\eqref{eq:SPG_armijo} and
the stopping map is $\|\mathsf{g}_1\|_\infty$ Eq.~\eqref{eq:gt_def}.
The repair map must therefore be the $\ell_2$ closest-point projection onto
$\Omega$ Eq.~\eqref{eq:occ_feasible} \cite{Calamai1987,Birgin2000},
\begin{equation}
P_w(\bx)=\arg\min_{\mathsf{n}\in\Omega}\ \tfrac{1}{2}\sum_{i\bk}(n_{i\bk}-x_{i\bk})^2.
\label{eq:app_Pw_variational}
\end{equation}

\emph{Clip-shift form.}
Expanding the constraints in Eq.~\eqref{eq:occ_feasible}, introduce a multiplier
$\lambda$ for $\sum_{i\bk} w_{\bk} n_{i\bk}=N_e$ and multipliers
$\mu_{i\bk},\nu_{i\bk}\ge 0$ for $n_{i\bk}\ge 0$ and $1-n_{i\bk}\ge 0$:
\begin{equation}
\mathcal{L}=\tfrac{1}{2}\sum_{i\bk}(n_{i\bk}-x_{i\bk})^2
+ \lambda\!\left(\sum_{i\bk} w_{\bk} n_{i\bk}-N_e\right)
- \sum_{i\bk}\mu_{i\bk}\, n_{i\bk}
- \sum_{i\bk}\nu_{i\bk}\,(1-n_{i\bk}).
\label{eq:app_Lagrangian}
\end{equation}
Stationarity $\partial\mathcal{L}/\partial n_{i\bk}=0$ gives
\begin{equation}
n_{i\bk}-x_{i\bk} + \lambda w_{\bk} - \mu_{i\bk} + \nu_{i\bk} = 0.
\label{eq:app_KKT}
\end{equation}
With $s_{i\bk}\equiv x_{i\bk}-\lambda w_{\bk}$ and complementarity
($\mu_{i\bk} n_{i\bk}=\nu_{i\bk}(1-n_{i\bk})=0$), each component satisfies
\begin{equation}
n_{i\bk}=
\begin{cases}
s_{i\bk}, & 0 \le s_{i\bk} \le 1,\\[4pt]
0, & s_{i\bk} < 0,\\[4pt]
1, & s_{i\bk} > 1,
\end{cases}
\label{eq:app_three_cases}
\end{equation}
at the physical endpoints $0$ and $1$ [Eq.~\eqref{eq:occ_constraint}].
This is Eq.~\eqref{eq:Pw}, with $\lambda$ fixed by Eq.~\eqref{eq:Gw}.
The factor $w_{\bk}$ in $s_{i\bk}$ is the Lagrange multiplier of the weighted
electron-count constraint, not a separate weighting of the projection metric.

\emph{Initial descent.}
Because $P_w$ is Euclidean projection onto convex $\Omega$, BMR
Lemma~2.1~\cite{Birgin2000} gives $\phi'(0)<0$ whenever $\bd_k\ne\mathsf{0}$,
as required by Eq.~\eqref{eq:SPG_armijo}.

\section{Power-functional regularisation}
\label{app:power_regularization}
For the separable power kernel $g(n)=n^{\alpha}$ with
$\alpha\in(\tfrac{1}{2},1)$, the derivative $g'(n)=\alpha n^{\alpha-1}$
diverges as $n\to 0^{+}$.
To obtain a finite coupling and occupation gradient for line searches, $g(n)$
is replaced below a floor $n_{\min}=\varepsilon$ by its tangent at
$n=\varepsilon$:
\begin{equation}
g(n)=
\begin{cases}
n^{\alpha}, & n \ge \varepsilon,\\[4pt]
\varepsilon^{\alpha} + \alpha\,\varepsilon^{\alpha-1}(n-\varepsilon),
& n < \varepsilon,
\end{cases}
\label{eq:app_power_g}
\end{equation}
with derivative
\begin{equation}
g'(n)=\alpha\,\max(n,\varepsilon)^{\alpha-1}.
\label{eq:app_power_gp}
\end{equation}
The piecewise definitions coincide at $n=\varepsilon$:
$g(\varepsilon)=\varepsilon^{\alpha}$ and
$g'(\varepsilon)=\alpha\,\varepsilon^{\alpha-1}$ from either branch, so $g$
and $g'$ are continuous at the join.
Equations~\eqref{eq:app_power_g} and~\eqref{eq:app_power_gp} enter the evaluation
of $E_{\xc}$ and its occupation derivatives on equal footing.
The default floor is $\varepsilon=10^{-8}$.


\begin{thebibliography}{43}%
  \makeatletter
  \providecommand \@ifxundefined [1]{%
   \@ifx{#1\undefined}
  }%
  \providecommand \@ifnum [1]{%
   \ifnum #1\expandafter \@firstoftwo
   \else \expandafter \@secondoftwo
   \fi
  }%
  \providecommand \@ifx [1]{%
   \ifx #1\expandafter \@firstoftwo
   \else \expandafter \@secondoftwo
   \fi
  }%
  \providecommand \natexlab [1]{#1}%
  \providecommand \enquote  [1]{``#1''}%
  \providecommand \bibnamefont  [1]{#1}%
  \providecommand \bibfnamefont [1]{#1}%
  \providecommand \citenamefont [1]{#1}%
  \providecommand \href@noop [0]{\@secondoftwo}%
  \providecommand \href [0]{\begingroup \@sanitize@url \@href}%
  \providecommand \@href[1]{\@@startlink{#1}\@@href}%
  \providecommand \@@href[1]{\endgroup#1\@@endlink}%
  \providecommand \@sanitize@url [0]{\catcode `\\12\catcode `\$12\catcode `\&12\catcode `\#12\catcode `\^12\catcode `\_12\catcode `\%12\relax}%
  \providecommand \@@startlink[1]{}%
  \providecommand \@@endlink[0]{}%
  \providecommand \url  [0]{\begingroup\@sanitize@url \@url }%
  \providecommand \@url [1]{\endgroup\@href {#1}{\urlprefix }}%
  \providecommand \urlprefix  [0]{URL }%
  \providecommand \Eprint [0]{\href }%
  \providecommand \doibase [0]{https://doi.org/}%
  \providecommand \selectlanguage [0]{\@gobble}%
  \providecommand \bibinfo  [0]{\@secondoftwo}%
  \providecommand \bibfield  [0]{\@secondoftwo}%
  \providecommand \translation [1]{[#1]}%
  \providecommand \BibitemOpen [0]{}%
  \providecommand \bibitemStop [0]{}%
  \providecommand \bibitemNoStop [0]{.\EOS\space}%
  \providecommand \EOS [0]{\spacefactor3000\relax}%
  \providecommand \BibitemShut  [1]{\csname bibitem#1\endcsname}%
  \let\auto@bib@innerbib\@empty
  \bibitem [{\citenamefont {Hohenberg}\ and\ \citenamefont {Kohn}(1964)}]{Hohenberg1964}%
    \BibitemOpen
    \bibfield  {author} {\bibinfo {author} {\bibfnamefont {P.}~\bibnamefont {Hohenberg}}\ and\ \bibinfo {author} {\bibfnamefont {W.}~\bibnamefont {Kohn}},\ }\bibfield  {title} {\bibinfo {title} {Inhomogeneous electron gas},\ }\href {https://doi.org/10.1103/PhysRev.136.B864} {\bibfield  {journal} {\bibinfo  {journal} {Phys. Rev.}\ }\textbf {\bibinfo {volume} {136}},\ \bibinfo {pages} {B864} (\bibinfo {year} {1964})}\BibitemShut {NoStop}%
  \bibitem [{\citenamefont {Kohn}\ and\ \citenamefont {Sham}(1965)}]{Kohn1965}%
    \BibitemOpen
    \bibfield  {author} {\bibinfo {author} {\bibfnamefont {W.}~\bibnamefont {Kohn}}\ and\ \bibinfo {author} {\bibfnamefont {L.~J.}\ \bibnamefont {Sham}},\ }\bibfield  {title} {\bibinfo {title} {Self-consistent equations including exchange and correlation effects},\ }\href {https://doi.org/10.1103/PhysRev.140.A1133} {\bibfield  {journal} {\bibinfo  {journal} {Phys. Rev.}\ }\textbf {\bibinfo {volume} {140}},\ \bibinfo {pages} {A1133} (\bibinfo {year} {1965})}\BibitemShut {NoStop}%
  \bibitem [{\citenamefont {Gilbert}(1975)}]{Gilbert1975}%
    \BibitemOpen
    \bibfield  {author} {\bibinfo {author} {\bibfnamefont {T.~L.}\ \bibnamefont {Gilbert}},\ }\bibfield  {title} {\bibinfo {title} {Hohenberg-kohn theorem for nonlocal external potentials},\ }\href {https://doi.org/10.1103/PhysRevB.12.2111} {\bibfield  {journal} {\bibinfo  {journal} {Phys. Rev. B}\ }\textbf {\bibinfo {volume} {12}},\ \bibinfo {pages} {2111} (\bibinfo {year} {1975})}\BibitemShut {NoStop}%
  \bibitem [{\citenamefont {Muller}(1984)}]{Muller1984}%
    \BibitemOpen
    \bibfield  {author} {\bibinfo {author} {\bibfnamefont {A.~M.~K.}\ \bibnamefont {Muller}},\ }\bibfield  {title} {\bibinfo {title} {Explicit approximate relation between reduced two- and one-particle density matrices},\ }\href {https://doi.org/10.1016/0375-9601(84)91034-X} {\bibfield  {journal} {\bibinfo  {journal} {Phys. Lett. A}\ }\textbf {\bibinfo {volume} {105}},\ \bibinfo {pages} {446} (\bibinfo {year} {1984})}\BibitemShut {NoStop}%
  \bibitem [{\citenamefont {Pernal}\ and\ \citenamefont {Giesbertz}(2016)}]{Pernal2016}%
    \BibitemOpen
    \bibfield  {author} {\bibinfo {author} {\bibfnamefont {K.}~\bibnamefont {Pernal}}\ and\ \bibinfo {author} {\bibfnamefont {K.~J.~H.}\ \bibnamefont {Giesbertz}},\ }\bibinfo {title} {Reduced density matrix functional theory (rdmft) and linear response time-dependent rdmft (td-rdmft)},\ in\ \href {https://doi.org/10.1007/128_2015_624} {\emph {\bibinfo {booktitle} {Density-Functional Methods for Excited States}}},\ \bibinfo {editor} {edited by\ \bibinfo {editor} {\bibfnamefont {N.}~\bibnamefont {Ferre}}, \bibinfo {editor} {\bibfnamefont {M.}~\bibnamefont {Filatov}},\ and\ \bibinfo {editor} {\bibfnamefont {M.}~\bibnamefont {Huix-Rotllant}}}\ (\bibinfo  {publisher} {Springer International Publishing},\ \bibinfo {address} {Cham},\ \bibinfo {year} {2016})\ pp.\ \bibinfo {pages} {125--183}\BibitemShut {NoStop}%
  \bibitem [{\citenamefont {BUIJSE}\ and\ \citenamefont {BAERENDS}(2002)}]{BB2002}%
    \BibitemOpen
    \bibfield  {author} {\bibinfo {author} {\bibfnamefont {M.~A.}\ \bibnamefont {BUIJSE}}\ and\ \bibinfo {author} {\bibfnamefont {E.~J.}\ \bibnamefont {BAERENDS}},\ }\bibfield  {title} {\bibinfo {title} {An approximate exchange-correlation hole density as a functional of the natural orbitals},\ }\href {https://doi.org/10.1080/00268970110070243} {\bibfield  {journal} {\bibinfo  {journal} {Molecular Physics}\ }\textbf {\bibinfo {volume} {100}},\ \bibinfo {pages} {401} (\bibinfo {year} {2002})},\ \Eprint {https://arxiv.org/abs/https://doi.org/10.1080/00268970110070243} {https://doi.org/10.1080/00268970110070243} \BibitemShut {NoStop}%
  \bibitem [{\citenamefont {Goedecker}\ and\ \citenamefont {Umrigar}(1998)}]{Goedecker1998}%
    \BibitemOpen
    \bibfield  {author} {\bibinfo {author} {\bibfnamefont {S.}~\bibnamefont {Goedecker}}\ and\ \bibinfo {author} {\bibfnamefont {C.~J.}\ \bibnamefont {Umrigar}},\ }\bibfield  {title} {\bibinfo {title} {Natural orbital functional for the many-electron problem},\ }\href {https://doi.org/10.1103/PhysRevLett.81.866} {\bibfield  {journal} {\bibinfo  {journal} {Phys. Rev. Lett.}\ }\textbf {\bibinfo {volume} {81}},\ \bibinfo {pages} {866} (\bibinfo {year} {1998})}\BibitemShut {NoStop}%
  \bibitem [{\citenamefont {Gritsenko}\ \emph {et~al.}(2005)\citenamefont {Gritsenko}, \citenamefont {Pernal},\ and\ \citenamefont {Baerends}}]{Gritsenko2005}%
    \BibitemOpen
    \bibfield  {author} {\bibinfo {author} {\bibfnamefont {O.}~\bibnamefont {Gritsenko}}, \bibinfo {author} {\bibfnamefont {K.}~\bibnamefont {Pernal}},\ and\ \bibinfo {author} {\bibfnamefont {E.~J.}\ \bibnamefont {Baerends}},\ }\bibfield  {title} {\bibinfo {title} {An improved density matrix functional by physically motivated repulsive corrections},\ }\href {https://doi.org/10.1063/1.1906203} {\bibfield  {journal} {\bibinfo  {journal} {The Journal of Chemical Physics}\ }\textbf {\bibinfo {volume} {122}},\ \bibinfo {pages} {204102} (\bibinfo {year} {2005})}\BibitemShut {NoStop}%
  \bibitem [{\citenamefont {Cs\'anyi}\ and\ \citenamefont {Arias}(2000)}]{Csanyi2000}%
    \BibitemOpen
    \bibfield  {author} {\bibinfo {author} {\bibfnamefont {G.}~\bibnamefont {Cs\'anyi}}\ and\ \bibinfo {author} {\bibfnamefont {T.~A.}\ \bibnamefont {Arias}},\ }\bibfield  {title} {\bibinfo {title} {Tensor product expansions for correlation in quantum many-body systems},\ }\href {https://doi.org/10.1103/PhysRevB.61.7348} {\bibfield  {journal} {\bibinfo  {journal} {Phys. Rev. B}\ }\textbf {\bibinfo {volume} {61}},\ \bibinfo {pages} {7348} (\bibinfo {year} {2000})}\BibitemShut {NoStop}%
  \bibitem [{\citenamefont {Cs\'anyi}\ \emph {et~al.}(2002)\citenamefont {Cs\'anyi}, \citenamefont {Goedecker},\ and\ \citenamefont {Arias}}]{Csanyi2002}%
    \BibitemOpen
    \bibfield  {author} {\bibinfo {author} {\bibfnamefont {G.}~\bibnamefont {Cs\'anyi}}, \bibinfo {author} {\bibfnamefont {S.}~\bibnamefont {Goedecker}},\ and\ \bibinfo {author} {\bibfnamefont {T.~A.}\ \bibnamefont {Arias}},\ }\bibfield  {title} {\bibinfo {title} {Improved tensor-product expansions for the two-particle density matrix},\ }\href {https://doi.org/10.1103/PhysRevA.65.032510} {\bibfield  {journal} {\bibinfo  {journal} {Phys. Rev. A}\ }\textbf {\bibinfo {volume} {65}},\ \bibinfo {pages} {032510} (\bibinfo {year} {2002})}\BibitemShut {NoStop}%
  \bibitem [{\citenamefont {Pernal}(2013)}]{Pernal2013}%
    \BibitemOpen
    \bibfield  {author} {\bibinfo {author} {\bibfnamefont {K.}~\bibnamefont {Pernal}},\ }\bibfield  {title} {\bibinfo {title} {The equivalence of the piris natural orbital functional 5 (PNOF5) and the antisymmetrized product of strongly orthogonal geminal theory},\ }\href {https://doi.org/https://doi.org/10.1016/j.comptc.2012.08.022} {\bibfield  {journal} {\bibinfo  {journal} {Computational and Theoretical Chemistry}\ }\textbf {\bibinfo {volume} {1003}},\ \bibinfo {pages} {127} (\bibinfo {year} {2013})},\ \bibinfo {note} {reduced Density Matrices: A Simpler Approach to Many-Electron Problems?}\BibitemShut {Stop}%
  \bibitem [{\citenamefont {Pernal}(2010)}]{Pernal2010}%
    \BibitemOpen
    \bibfield  {author} {\bibinfo {author} {\bibfnamefont {K.}~\bibnamefont {Pernal}},\ }\bibfield  {title} {\bibinfo {title} {Long-range density-matrix-functional theory: Application to a modified homogeneous electron gas},\ }\href {https://doi.org/10.1103/PhysRevA.81.052511} {\bibfield  {journal} {\bibinfo  {journal} {Phys. Rev. A}\ }\textbf {\bibinfo {volume} {81}},\ \bibinfo {pages} {052511} (\bibinfo {year} {2010})}\BibitemShut {NoStop}%
  \bibitem [{\citenamefont {Ai}\ \emph {et~al.}(2023)\citenamefont {Ai}, \citenamefont {Su},\ and\ \citenamefont {Fang}}]{Ai2023}%
    \BibitemOpen
    \bibfield  {author} {\bibinfo {author} {\bibfnamefont {W.}~\bibnamefont {Ai}}, \bibinfo {author} {\bibfnamefont {N.~Q.}\ \bibnamefont {Su}},\ and\ \bibinfo {author} {\bibfnamefont {W.-H.}\ \bibnamefont {Fang}},\ }\bibfield  {title} {\bibinfo {title} {Short-range screened density matrix functional for proper descriptions of thermochemistry, thermochemical kinetics, nonbonded interactions, and singlet diradicals},\ }\href {https://doi.org/10.1063/5.0169234} {\bibfield  {journal} {\bibinfo  {journal} {The Journal of Chemical Physics}\ }\textbf {\bibinfo {volume} {159}},\ \bibinfo {pages} {174110} (\bibinfo {year} {2023})}\BibitemShut {NoStop}%
  \bibitem [{\citenamefont {Sharma}\ \emph {et~al.}(2008)\citenamefont {Sharma}, \citenamefont {Dewhurst}, \citenamefont {Lathiotakis},\ and\ \citenamefont {Gross}}]{Sharma2008}%
    \BibitemOpen
    \bibfield  {author} {\bibinfo {author} {\bibfnamefont {S.}~\bibnamefont {Sharma}}, \bibinfo {author} {\bibfnamefont {J.~K.}\ \bibnamefont {Dewhurst}}, \bibinfo {author} {\bibfnamefont {N.~N.}\ \bibnamefont {Lathiotakis}},\ and\ \bibinfo {author} {\bibfnamefont {E.~K.~U.}\ \bibnamefont {Gross}},\ }\bibfield  {title} {\bibinfo {title} {Reduced density matrix functional for many-electron systems},\ }\href {https://doi.org/10.1103/PhysRevB.78.201103} {\bibfield  {journal} {\bibinfo  {journal} {Phys. Rev. B}\ }\textbf {\bibinfo {volume} {78}},\ \bibinfo {pages} {201103} (\bibinfo {year} {2008})}\BibitemShut {NoStop}%
  \bibitem [{\citenamefont {Baldsiefen}\ \emph {et~al.}(2017)\citenamefont {Baldsiefen}, \citenamefont {Cangi}, \citenamefont {Eich},\ and\ \citenamefont {Gross}}]{Baldsiefen2017}%
    \BibitemOpen
    \bibfield  {author} {\bibinfo {author} {\bibfnamefont {T.}~\bibnamefont {Baldsiefen}}, \bibinfo {author} {\bibfnamefont {A.}~\bibnamefont {Cangi}}, \bibinfo {author} {\bibfnamefont {F.~G.}\ \bibnamefont {Eich}},\ and\ \bibinfo {author} {\bibfnamefont {E.~K.~U.}\ \bibnamefont {Gross}},\ }\bibfield  {title} {\bibinfo {title} {Exchange-correlation approximations for reduced-density-matrix-functional theory at finite temperature: Capturing magnetic phase transitions in the homogeneous electron gas},\ }\href {https://doi.org/10.1103/PhysRevA.96.062508} {\bibfield  {journal} {\bibinfo  {journal} {Phys. Rev. A}\ }\textbf {\bibinfo {volume} {96}},\ \bibinfo {pages} {062508} (\bibinfo {year} {2017})}\BibitemShut {NoStop}%
  \bibitem [{\citenamefont {Vladaj}\ \emph {et~al.}(2024)\citenamefont {Vladaj}, \citenamefont {Marécat}, \citenamefont {Senjean},\ and\ \citenamefont {Saubanère}}]{Vladaj2024}%
    \BibitemOpen
    \bibfield  {author} {\bibinfo {author} {\bibfnamefont {M.}~\bibnamefont {Vladaj}}, \bibinfo {author} {\bibfnamefont {Q.}~\bibnamefont {Marécat}}, \bibinfo {author} {\bibfnamefont {B.}~\bibnamefont {Senjean}},\ and\ \bibinfo {author} {\bibfnamefont {M.}~\bibnamefont {Saubanère}},\ }\bibfield  {title} {\bibinfo {title} {Variational minimization scheme for the one-particle reduced density matrix functional theory in the ensemble n-representability domain},\ }\href {https://doi.org/10.1063/5.0219898} {\bibfield  {journal} {\bibinfo  {journal} {The Journal of Chemical Physics}\ }\textbf {\bibinfo {volume} {161}},\ \bibinfo {pages} {074105} (\bibinfo {year} {2024})}\BibitemShut {NoStop}%
  \bibitem [{\citenamefont {Piris}\ and\ \citenamefont {Mitxelena}(2021)}]{DoNOF2021}%
    \BibitemOpen
    \bibfield  {author} {\bibinfo {author} {\bibfnamefont {M.}~\bibnamefont {Piris}}\ and\ \bibinfo {author} {\bibfnamefont {I.}~\bibnamefont {Mitxelena}},\ }\bibfield  {title} {\bibinfo {title} {Donof: An open-source implementation of natural-orbital-functional-based methods for quantum chemistry},\ }\href {https://doi.org/10.1016/j.cpc.2020.107651} {\bibfield  {journal} {\bibinfo  {journal} {Comput. Phys. Commun.}\ }\textbf {\bibinfo {volume} {259}},\ \bibinfo {pages} {107651} (\bibinfo {year} {2021})}\BibitemShut {NoStop}%
  \bibitem [{\citenamefont {Lew-Yee}\ \emph {et~al.}(2026)\citenamefont {Lew-Yee}, \citenamefont {Mitxelena}, \citenamefont {del Campo},\ and\ \citenamefont {Piris}}]{DoNOF2026}%
    \BibitemOpen
    \bibfield  {author} {\bibinfo {author} {\bibfnamefont {J.~F.~H.}\ \bibnamefont {Lew-Yee}}, \bibinfo {author} {\bibfnamefont {I.}~\bibnamefont {Mitxelena}}, \bibinfo {author} {\bibfnamefont {J.~M.}\ \bibnamefont {del Campo}},\ and\ \bibinfo {author} {\bibfnamefont {M.}~\bibnamefont {Piris}},\ }\bibfield  {title} {\bibinfo {title} {Donof 2.0: A modern open-source electronic structure program for natural orbital functionals},\ }\href {https://doi.org/10.1063/5.0316927} {\bibfield  {journal} {\bibinfo  {journal} {The Journal of Chemical Physics}\ }\textbf {\bibinfo {volume} {164}},\ \bibinfo {pages} {072501} (\bibinfo {year} {2026})}\BibitemShut {NoStop}%
  \bibitem [{elk()}]{elk}%
    \BibitemOpen
    \href@noop {} {\bibinfo {title} {{The Elk Code}}},\ \bibinfo {howpublished} {\url{http://elk.sourceforge.net/}}\BibitemShut {NoStop}%
  \bibitem [{\citenamefont {Giannozzi}\ \emph {et~al.}(2009)\citenamefont {Giannozzi}, \citenamefont {Baroni}, \citenamefont {Bonini}, \citenamefont {Calandra}, \citenamefont {Car}, \citenamefont {Cavazzoni}, \citenamefont {Ceresoli}, \citenamefont {Chiarotti}, \citenamefont {Cococcioni}, \citenamefont {Dabo}, \citenamefont {Dal~Corso}, \citenamefont {de~Gironcoli}, \citenamefont {Fabris}, \citenamefont {Fratesi}, \citenamefont {Gebauer}, \citenamefont {Gerstmann}, \citenamefont {Gougoussis}, \citenamefont {Kokalj}, \citenamefont {Lazzeri}, \citenamefont {Martin-Samos}, \citenamefont {Marzari}, \citenamefont {Mauri}, \citenamefont {Mazzarello}, \citenamefont {Paolini}, \citenamefont {Pasquarello}, \citenamefont {Paulatto}, \citenamefont {Sbraccia}, \citenamefont {Scandolo}, \citenamefont {Sclauzero}, \citenamefont {Seitsonen}, \citenamefont {Smogunov}, \citenamefont {Umari},\ and\ \citenamefont {Wentzcovitch}}]{Giannozzi2009}%
    \BibitemOpen
    \bibfield  {author} {\bibinfo {author} {\bibfnamefont {P.}~\bibnamefont {Giannozzi}}, \bibinfo {author} {\bibfnamefont {S.}~\bibnamefont {Baroni}}, \bibinfo {author} {\bibfnamefont {N.}~\bibnamefont {Bonini}}, \bibinfo {author} {\bibfnamefont {M.}~\bibnamefont {Calandra}}, \bibinfo {author} {\bibfnamefont {R.}~\bibnamefont {Car}}, \bibinfo {author} {\bibfnamefont {C.}~\bibnamefont {Cavazzoni}}, \bibinfo {author} {\bibfnamefont {D.}~\bibnamefont {Ceresoli}}, \bibinfo {author} {\bibfnamefont {G.~L.}\ \bibnamefont {Chiarotti}}, \bibinfo {author} {\bibfnamefont {M.}~\bibnamefont {Cococcioni}}, \bibinfo {author} {\bibfnamefont {I.}~\bibnamefont {Dabo}}, \bibinfo {author} {\bibfnamefont {A.}~\bibnamefont {Dal~Corso}}, \bibinfo {author} {\bibfnamefont {S.}~\bibnamefont {de~Gironcoli}}, \bibinfo {author} {\bibfnamefont {S.}~\bibnamefont {Fabris}}, \bibinfo {author} {\bibfnamefont {G.}~\bibnamefont {Fratesi}}, \bibinfo {author} {\bibfnamefont {R.}~\bibnamefont {Gebauer}}, \bibinfo {author} {\bibfnamefont {U.}~\bibnamefont {Gerstmann}}, \bibinfo {author} {\bibfnamefont {C.}~\bibnamefont {Gougoussis}}, \bibinfo {author} {\bibfnamefont {A.}~\bibnamefont {Kokalj}}, \bibinfo {author} {\bibfnamefont {M.}~\bibnamefont {Lazzeri}}, \bibinfo {author} {\bibfnamefont {L.}~\bibnamefont {Martin-Samos}}, \bibinfo {author} {\bibfnamefont {N.}~\bibnamefont {Marzari}}, \bibinfo {author} {\bibfnamefont {F.}~\bibnamefont {Mauri}}, \bibinfo {author} {\bibfnamefont {R.}~\bibnamefont {Mazzarello}}, \bibinfo {author} {\bibfnamefont {S.}~\bibnamefont {Paolini}}, \bibinfo {author} {\bibfnamefont {A.}~\bibnamefont {Pasquarello}}, \bibinfo {author} {\bibfnamefont {L.}~\bibnamefont {Paulatto}}, \bibinfo {author} {\bibfnamefont {C.}~\bibnamefont {Sbraccia}}, \bibinfo {author} {\bibfnamefont {S.}~\bibnamefont {Scandolo}}, \bibinfo {author} {\bibfnamefont {G.}~\bibnamefont {Sclauzero}}, \bibinfo {author} {\bibfnamefont {A.~P.}\ \bibnamefont {Seitsonen}}, \bibinfo {author} {\bibfnamefont {A.}~\bibnamefont {Smogunov}}, \bibinfo {author} {\bibfnamefont {P.}~\bibnamefont {Umari}},\ and\ \bibinfo {author} {\bibfnamefont {R.~M.}\ \bibnamefont {Wentzcovitch}},\ }\bibfield  {title} {\bibinfo {title} {Quantum espresso: a modular and open-source software project for quantum simulations of materials},\ }\href {https://doi.org/10.1088/0953-8984/21/39/395502} {\bibfield  {journal} {\bibinfo  {journal} {Journal of Physics: Condensed Matter}\ }\textbf {\bibinfo {volume} {21}},\ \bibinfo {pages} {395502} (\bibinfo {year} {2009})}\BibitemShut {NoStop}%
  \bibitem [{\citenamefont {Luo}\ \emph {et~al.}(2025)\citenamefont {Luo}, \citenamefont {Wang},\ and\ \citenamefont {Ren}}]{Luo2025}%
    \BibitemOpen
    \bibfield  {author} {\bibinfo {author} {\bibfnamefont {K.}~\bibnamefont {Luo}}, \bibinfo {author} {\bibfnamefont {T.}~\bibnamefont {Wang}},\ and\ \bibinfo {author} {\bibfnamefont {X.}~\bibnamefont {Ren}},\ }\bibfield  {title} {\bibinfo {title} {Direct minimization on the complex stiefel manifold in kohn-sham density functional theory for finite and extended systems},\ }\href {https://doi.org/https://doi.org/10.1016/j.cpc.2025.109596} {\bibfield  {journal} {\bibinfo  {journal} {Computer Physics Communications}\ }\textbf {\bibinfo {volume} {312}},\ \bibinfo {pages} {109596} (\bibinfo {year} {2025})}\BibitemShut {NoStop}%
  \bibitem [{\citenamefont {Lin}(2016)}]{Lin2016}%
    \BibitemOpen
    \bibfield  {author} {\bibinfo {author} {\bibfnamefont {L.}~\bibnamefont {Lin}},\ }\bibfield  {title} {\bibinfo {title} {Adaptively compressed exchange operator},\ }\href {https://doi.org/10.1021/acs.jctc.6b00092} {\bibfield  {journal} {\bibinfo  {journal} {Journal of Chemical Theory and Computation}\ }\textbf {\bibinfo {volume} {12}},\ \bibinfo {pages} {2242} (\bibinfo {year} {2016})},\ \bibinfo {note} {pMID: 27045571},\ \Eprint {https://arxiv.org/abs/https://doi.org/10.1021/acs.jctc.6b00092} {https://doi.org/10.1021/acs.jctc.6b00092} \BibitemShut {NoStop}%
  \bibitem [{\citenamefont {L\"owdin}(1955)}]{Lowdin1955}%
    \BibitemOpen
    \bibfield  {author} {\bibinfo {author} {\bibfnamefont {P.-O.}\ \bibnamefont {L\"owdin}},\ }\bibfield  {title} {\bibinfo {title} {Quantum theory of many-particle systems. i. physical interpretations by means of density matrices, natural spin-orbitals, and convergence problems in the method of configurational interaction},\ }\href {https://doi.org/10.1103/PhysRev.97.1474} {\bibfield  {journal} {\bibinfo  {journal} {Phys. Rev.}\ }\textbf {\bibinfo {volume} {97}},\ \bibinfo {pages} {1474} (\bibinfo {year} {1955})}\BibitemShut {NoStop}%
  \bibitem [{\citenamefont {Coleman}(1961)}]{Coleman1961}%
    \BibitemOpen
    \bibfield  {author} {\bibinfo {author} {\bibfnamefont {A.~J.}\ \bibnamefont {Coleman}},\ }\bibfield  {title} {\bibinfo {title} {Density matrices of n-fermion systems},\ }\href {https://doi.org/10.4153/CMB-1961-023-7} {\bibfield  {journal} {\bibinfo  {journal} {Canadian Mathematical Bulletin}\ }\textbf {\bibinfo {volume} {4}},\ \bibinfo {pages} {209–212} (\bibinfo {year} {1961})}\BibitemShut {NoStop}%
  \bibitem [{\citenamefont {Coleman}(1963)}]{Coleman1963}%
    \BibitemOpen
    \bibfield  {author} {\bibinfo {author} {\bibfnamefont {A.~J.}\ \bibnamefont {Coleman}},\ }\bibfield  {title} {\bibinfo {title} {Structure of fermion density matrices},\ }\href {https://doi.org/10.1103/RevModPhys.35.668} {\bibfield  {journal} {\bibinfo  {journal} {Rev. Mod. Phys.}\ }\textbf {\bibinfo {volume} {35}},\ \bibinfo {pages} {668} (\bibinfo {year} {1963})}\BibitemShut {NoStop}%
  \bibitem [{\citenamefont {Gygi}\ and\ \citenamefont {Baldereschi}(1986)}]{Gygi1986}%
    \BibitemOpen
    \bibfield  {author} {\bibinfo {author} {\bibfnamefont {F.}~\bibnamefont {Gygi}}\ and\ \bibinfo {author} {\bibfnamefont {A.}~\bibnamefont {Baldereschi}},\ }\bibfield  {title} {\bibinfo {title} {Self-consistent hartree-fock and screened-exchange calculations in solids: Application to silicon},\ }\href {https://doi.org/10.1103/PhysRevB.34.4405} {\bibfield  {journal} {\bibinfo  {journal} {Phys. Rev. B}\ }\textbf {\bibinfo {volume} {34}},\ \bibinfo {pages} {4405(R)} (\bibinfo {year} {1986})}\BibitemShut {NoStop}%
  \bibitem [{\citenamefont {Kuhn}\ and\ \citenamefont {Tucker}(2014)}]{Kuhn2014}%
    \BibitemOpen
    \bibfield  {author} {\bibinfo {author} {\bibfnamefont {H.~W.}\ \bibnamefont {Kuhn}}\ and\ \bibinfo {author} {\bibfnamefont {A.~W.}\ \bibnamefont {Tucker}},\ }\bibinfo {title} {Nonlinear programming},\ in\ \href {https://doi.org/10.1007/978-3-0348-0439-4_11} {\emph {\bibinfo {booktitle} {Traces and Emergence of Nonlinear Programming}}},\ \bibinfo {editor} {edited by\ \bibinfo {editor} {\bibfnamefont {G.}~\bibnamefont {Giorgi}}\ and\ \bibinfo {editor} {\bibfnamefont {T.~H.}\ \bibnamefont {Kjeldsen}}}\ (\bibinfo  {publisher} {Springer Basel},\ \bibinfo {address} {Basel},\ \bibinfo {year} {2014})\ pp.\ \bibinfo {pages} {247--258}\BibitemShut {NoStop}%
  \bibitem [{\citenamefont {Yao}\ \emph {et~al.}(2021)\citenamefont {Yao}, \citenamefont {Fang},\ and\ \citenamefont {Su}}]{Yao2021}%
    \BibitemOpen
    \bibfield  {author} {\bibinfo {author} {\bibfnamefont {Y.~F.}\ \bibnamefont {Yao}}, \bibinfo {author} {\bibfnamefont {W.~H.}\ \bibnamefont {Fang}},\ and\ \bibinfo {author} {\bibfnamefont {N.~Q.}\ \bibnamefont {Su}},\ }\bibfield  {title} {\bibinfo {title} {Handling ensemble n-representability constraint in explicit-by-implicit manner},\ }\href {https://doi.org/10.1021/acs.jpclett.1c01835} {\bibfield  {journal} {\bibinfo  {journal} {J. Phys. Chem. Lett.}\ }\textbf {\bibinfo {volume} {12}},\ \bibinfo {pages} {6788} (\bibinfo {year} {2021})}\BibitemShut {NoStop}%
  \bibitem [{\citenamefont {Yao}\ \emph {et~al.}(2022)\citenamefont {Yao}, \citenamefont {Zhang}, \citenamefont {Fang},\ and\ \citenamefont {Su}}]{Yao2022}%
    \BibitemOpen
    \bibfield  {author} {\bibinfo {author} {\bibfnamefont {Y.-F.}\ \bibnamefont {Yao}}, \bibinfo {author} {\bibfnamefont {Z.}~\bibnamefont {Zhang}}, \bibinfo {author} {\bibfnamefont {W.-H.}\ \bibnamefont {Fang}},\ and\ \bibinfo {author} {\bibfnamefont {N.~Q.}\ \bibnamefont {Su}},\ }\bibfield  {title} {\bibinfo {title} {Explicit-by-implicit treatment of natural orbital occupations using first- and second-order optimization algorithms: A comparative study},\ }\href {https://doi.org/10.1021/acs.jpca.2c02345} {\bibfield  {journal} {\bibinfo  {journal} {The Journal of Physical Chemistry A}\ }\textbf {\bibinfo {volume} {126}},\ \bibinfo {pages} {5654} (\bibinfo {year} {2022})},\ \bibinfo {note} {pMID: 35950981},\ \Eprint {https://arxiv.org/abs/https://doi.org/10.1021/acs.jpca.2c02345} {https://doi.org/10.1021/acs.jpca.2c02345} \BibitemShut {NoStop}%
  \bibitem [{\citenamefont {Birgin}\ \emph {et~al.}(2000)\citenamefont {Birgin}, \citenamefont {Mart\'{\i}nez},\ and\ \citenamefont {Raydan}}]{Birgin2000}%
    \BibitemOpen
    \bibfield  {author} {\bibinfo {author} {\bibfnamefont {E.~G.}\ \bibnamefont {Birgin}}, \bibinfo {author} {\bibfnamefont {J.~M.}\ \bibnamefont {Mart\'{\i}nez}},\ and\ \bibinfo {author} {\bibfnamefont {M.}~\bibnamefont {Raydan}},\ }\bibfield  {title} {\bibinfo {title} {Nonmonotone spectral projected gradient methods on convex sets},\ }\href {https://doi.org/10.1137/S1052623497330963} {\bibfield  {journal} {\bibinfo  {journal} {SIAM Journal on Optimization}\ }\textbf {\bibinfo {volume} {10}},\ \bibinfo {pages} {1196} (\bibinfo {year} {2000})},\ \Eprint {https://arxiv.org/abs/https://doi.org/10.1137/S1052623497330963} {https://doi.org/10.1137/S1052623497330963} \BibitemShut {NoStop}%
  \bibitem [{\citenamefont {Calamai}\ and\ \citenamefont {Mor\'{e}}(1987)}]{Calamai1987}%
    \BibitemOpen
    \bibfield  {author} {\bibinfo {author} {\bibfnamefont {P.~H.}\ \bibnamefont {Calamai}}\ and\ \bibinfo {author} {\bibfnamefont {J.~J.}\ \bibnamefont {Mor\'{e}}},\ }\bibfield  {title} {\bibinfo {title} {Projected gradient methods for linearly constrained problems},\ }\href {https://doi.org/10.1007/BF02592073} {\bibfield  {journal} {\bibinfo  {journal} {Mathematical Programming}\ }\textbf {\bibinfo {volume} {39}},\ \bibinfo {pages} {93} (\bibinfo {year} {1987})}\BibitemShut {NoStop}%
  \bibitem [{\citenamefont {Barzilai}\ and\ \citenamefont {Borwein}(1988)}]{Barzilai1988}%
    \BibitemOpen
    \bibfield  {author} {\bibinfo {author} {\bibfnamefont {J.}~\bibnamefont {Barzilai}}\ and\ \bibinfo {author} {\bibfnamefont {J.~M.}\ \bibnamefont {Borwein}},\ }\bibfield  {title} {\bibinfo {title} {Two-point step size gradient methods},\ }\href {https://doi.org/10.1093/imanum/8.1.141} {\bibfield  {journal} {\bibinfo  {journal} {IMA Journal of Numerical Analysis}\ }\textbf {\bibinfo {volume} {8}},\ \bibinfo {pages} {141} (\bibinfo {year} {1988})}\BibitemShut {NoStop}%
  \bibitem [{\citenamefont {Armijo}(1966)}]{Armijo1966}%
    \BibitemOpen
    \bibfield  {author} {\bibinfo {author} {\bibfnamefont {L.}~\bibnamefont {Armijo}},\ }\bibfield  {title} {\bibinfo {title} {Minimization of functions having {L}ipschitz continuous first partial derivatives},\ }\href {https://doi.org/10.2140/pjm.1966.16.1} {\bibfield  {journal} {\bibinfo  {journal} {Pacific Journal of Mathematics}\ }\textbf {\bibinfo {volume} {16}},\ \bibinfo {pages} {1} (\bibinfo {year} {1966})}\BibitemShut {NoStop}%
  \bibitem [{\citenamefont {Grippo}\ \emph {et~al.}(1986)\citenamefont {Grippo}, \citenamefont {Lampariello},\ and\ \citenamefont {Lucidi}}]{Grippo1986}%
    \BibitemOpen
    \bibfield  {author} {\bibinfo {author} {\bibfnamefont {L.}~\bibnamefont {Grippo}}, \bibinfo {author} {\bibfnamefont {F.}~\bibnamefont {Lampariello}},\ and\ \bibinfo {author} {\bibfnamefont {S.}~\bibnamefont {Lucidi}},\ }\bibfield  {title} {\bibinfo {title} {A nonmonotone line search technique for {N}ewton's method},\ }\href {https://doi.org/10.1137/0723046} {\bibfield  {journal} {\bibinfo  {journal} {SIAM Journal on Numerical Analysis}\ }\textbf {\bibinfo {volume} {23}},\ \bibinfo {pages} {707} (\bibinfo {year} {1986})}\BibitemShut {NoStop}%
  \bibitem [{\citenamefont {Schlipf}\ and\ \citenamefont {Gygi}(2015)}]{Schlipf2015}%
    \BibitemOpen
    \bibfield  {author} {\bibinfo {author} {\bibfnamefont {M.}~\bibnamefont {Schlipf}}\ and\ \bibinfo {author} {\bibfnamefont {F.}~\bibnamefont {Gygi}},\ }\bibfield  {title} {\bibinfo {title} {Optimization algorithm for the generation of oncv pseudopotentials},\ }\href {https://doi.org/https://doi.org/10.1016/j.cpc.2015.05.011} {\bibfield  {journal} {\bibinfo  {journal} {Computer Physics Communications}\ }\textbf {\bibinfo {volume} {196}},\ \bibinfo {pages} {36} (\bibinfo {year} {2015})}\BibitemShut {NoStop}%
  \bibitem [{\citenamefont {Martyna}\ and\ \citenamefont {Tuckerman}(1999)}]{Martyna1999}%
    \BibitemOpen
    \bibfield  {author} {\bibinfo {author} {\bibfnamefont {G.~J.}\ \bibnamefont {Martyna}}\ and\ \bibinfo {author} {\bibfnamefont {M.~E.}\ \bibnamefont {Tuckerman}},\ }\bibfield  {title} {\bibinfo {title} {A reciprocal space based method for treating long range interactions in ab initio and force-field-based calculations in clusters},\ }\href {https://doi.org/10.1063/1.477923} {\bibfield  {journal} {\bibinfo  {journal} {The Journal of Chemical Physics}\ }\textbf {\bibinfo {volume} {110}},\ \bibinfo {pages} {2810} (\bibinfo {year} {1999})}\BibitemShut {NoStop}%
  \bibitem [{\citenamefont {Cartier}\ and\ \citenamefont {Giesbertz}(2025)}]{Cartier2025}%
    \BibitemOpen
    \bibfield  {author} {\bibinfo {author} {\bibfnamefont {N.~G.}\ \bibnamefont {Cartier}}\ and\ \bibinfo {author} {\bibfnamefont {K.~J.~H.}\ \bibnamefont {Giesbertz}},\ }\bibfield  {title} {\bibinfo {title} {Impact of parametrizations of the one-body reduced density matrix on the energy landscape},\ }\href {https://doi.org/10.1021/acs.jpclett.5c00308} {\bibfield  {journal} {\bibinfo  {journal} {The Journal of Physical Chemistry Letters}\ }\textbf {\bibinfo {volume} {16}},\ \bibinfo {pages} {3822} (\bibinfo {year} {2025})},\ \Eprint {https://arxiv.org/abs/https://pubs.acs.org/jpclcd/article-pdf/16/15/3822/42039217/jz5c00308.pdf} {https://pubs.acs.org/jpclcd/article-pdf/16/15/3822/42039217/jz5c00308.pdf} \BibitemShut {NoStop}%
  \bibitem [{\citenamefont {Mori-S\'anchez}\ \emph {et~al.}(2008)\citenamefont {Mori-S\'anchez}, \citenamefont {Cohen},\ and\ \citenamefont {Yang}}]{Mori-Sanchez2008}%
    \BibitemOpen
    \bibfield  {author} {\bibinfo {author} {\bibfnamefont {P.}~\bibnamefont {Mori-S\'anchez}}, \bibinfo {author} {\bibfnamefont {A.~J.}\ \bibnamefont {Cohen}},\ and\ \bibinfo {author} {\bibfnamefont {W.}~\bibnamefont {Yang}},\ }\bibfield  {title} {\bibinfo {title} {Localization and delocalization errors in density functional theory and implications for band-gap prediction},\ }\href {https://doi.org/10.1103/PhysRevLett.100.146401} {\bibfield  {journal} {\bibinfo  {journal} {Phys. Rev. Lett.}\ }\textbf {\bibinfo {volume} {100}},\ \bibinfo {pages} {146401} (\bibinfo {year} {2008})}\BibitemShut {NoStop}%
  \bibitem [{\citenamefont {Li}\ and\ \citenamefont {Yang}(2017)}]{Li2017}%
    \BibitemOpen
    \bibfield  {author} {\bibinfo {author} {\bibfnamefont {C.}~\bibnamefont {Li}}\ and\ \bibinfo {author} {\bibfnamefont {W.}~\bibnamefont {Yang}},\ }\bibfield  {title} {\bibinfo {title} {On the piecewise convex or concave nature of ground state energy as a function of fractional number of electrons for approximate density functionals},\ }\href {https://doi.org/10.1063/1.4974988} {\bibfield  {journal} {\bibinfo  {journal} {The Journal of Chemical Physics}\ }\textbf {\bibinfo {volume} {146}},\ \bibinfo {pages} {074107} (\bibinfo {year} {2017})}\BibitemShut {NoStop}%
  \bibitem [{\citenamefont {Perdew}\ \emph {et~al.}(1996)\citenamefont {Perdew}, \citenamefont {Burke},\ and\ \citenamefont {Ernzerhof}}]{PBE1996}%
    \BibitemOpen
    \bibfield  {author} {\bibinfo {author} {\bibfnamefont {J.~P.}\ \bibnamefont {Perdew}}, \bibinfo {author} {\bibfnamefont {K.}~\bibnamefont {Burke}},\ and\ \bibinfo {author} {\bibfnamefont {M.}~\bibnamefont {Ernzerhof}},\ }\bibfield  {title} {\bibinfo {title} {Generalized gradient approximation made simple},\ }\href {https://doi.org/10.1103/PhysRevLett.77.3865} {\bibfield  {journal} {\bibinfo  {journal} {Phys. Rev. Lett.}\ }\textbf {\bibinfo {volume} {77}},\ \bibinfo {pages} {3865} (\bibinfo {year} {1996})}\BibitemShut {NoStop}%
  \bibitem [{\citenamefont {Wang}\ and\ \citenamefont {Baerends}(2022)}]{Wang2022}%
    \BibitemOpen
    \bibfield  {author} {\bibinfo {author} {\bibfnamefont {J.}~\bibnamefont {Wang}}\ and\ \bibinfo {author} {\bibfnamefont {E.~J.}\ \bibnamefont {Baerends}},\ }\bibfield  {title} {\bibinfo {title} {Self-consistent-field method for correlated many-electron systems with an entropic cumulant energy},\ }\href {https://doi.org/10.1103/PhysRevLett.128.013001} {\bibfield  {journal} {\bibinfo  {journal} {Phys. Rev. Lett.}\ }\textbf {\bibinfo {volume} {128}},\ \bibinfo {pages} {013001} (\bibinfo {year} {2022})}\BibitemShut {NoStop}%
  \bibitem [{\citenamefont {McSkimin}\ and\ \citenamefont {Andreatch}(1964)}]{McSkimin1964}%
    \BibitemOpen
    \bibfield  {author} {\bibinfo {author} {\bibfnamefont {H.~J.}\ \bibnamefont {McSkimin}}\ and\ \bibinfo {author} {\bibfnamefont {P.}~\bibnamefont {Andreatch}},\ }\bibfield  {title} {\bibinfo {title} {Elastic moduli of silicon vs hydrostatic pressure at {25.0\ifmmode^\circ\else\textdegree\fi C} and $-{195.8}\ifmmode^\circ\else\textdegree\fi c$},\ }\href {https://doi.org/10.1063/1.1713799} {\bibfield  {journal} {\bibinfo  {journal} {J. Appl. Phys.}\ }\textbf {\bibinfo {volume} {35}},\ \bibinfo {pages} {2161} (\bibinfo {year} {1964})}\BibitemShut {NoStop}%
  \bibitem [{\citenamefont {Sharma}\ \emph {et~al.}(2013)\citenamefont {Sharma}, \citenamefont {Dewhurst}, \citenamefont {Shallcross},\ and\ \citenamefont {Gross}}]{Sharma2013}%
    \BibitemOpen
    \bibfield  {author} {\bibinfo {author} {\bibfnamefont {S.}~\bibnamefont {Sharma}}, \bibinfo {author} {\bibfnamefont {J.~K.}\ \bibnamefont {Dewhurst}}, \bibinfo {author} {\bibfnamefont {S.}~\bibnamefont {Shallcross}},\ and\ \bibinfo {author} {\bibfnamefont {E.~K.~U.}\ \bibnamefont {Gross}},\ }\bibfield  {title} {\bibinfo {title} {Spectral density and metal-insulator phase transition in mott insulators within reduced density matrix functional theory},\ }\href {https://doi.org/10.1103/PhysRevLett.110.116403} {\bibfield  {journal} {\bibinfo  {journal} {Phys. Rev. Lett.}\ }\textbf {\bibinfo {volume} {110}},\ \bibinfo {pages} {116403} (\bibinfo {year} {2013})}\BibitemShut {NoStop}%
  \end{thebibliography}
%
  
\end{document}